\documentclass[letterpaper,twoside,twocolumn,english,reprint, aps, prb, superscriptaddress]{revtex4-1}
\usepackage{amsmath}
\usepackage{amssymb}

\usepackage{ifthen}
\usepackage{iftex}

\newcommand{\SetupXeLuaTex}{%
	\usepackage{fontspec}
	\setmainfont[Ligatures=TeX]{Linux Libertine O}
	\setsansfont[Ligatures=TeX]{Linux Biolinum O}
	\setmonofont{Linux Libertine Mono O}
	\usepackage{unicode-math}
	\setmathfont{XITS Math}
	\setmathfont[range=\mathup/{num,latin,Latin,greek,Greek}]{Linux Libertine O}
	\setmathfont[range=\mathit/{num,latin,Latin,greek,Greek}]{Linux Libertine O Italic}
	\setmathfont[range=\mathbfup/{num,latin,Latin,greek,Greek}]{Linux Libertine O Bold}
	\setmathfont[range=\mathbfit/{num,latin,Latin,greek,Greek}]{Linux Libertine O Bold Italic}
	\renewcommand{\slshape}{\itshape}
    \usepackage{microtype}
	
	\usepackage{polyglossia}
	\setdefaultlanguage[variant=american]{english}

}

\newcommand{\SetupPDFLaTeX}{%
	\usepackage[english]{babel}
}

\ifXeTeX 
    \SetupXeLuaTex
\else
    \ifLuaTeX
        \SetupXeLuaTex
    \else
       \SetupPDFLaTeX
    \fi
\fi

\newcommand{\noun}[1]{\textsc{#1}}

\usepackage{color}
\usepackage{units}
\usepackage{graphicx}
\usepackage{esint}

\usepackage[unicode=true,pdfusetitle,
 bookmarks=true,bookmarksnumbered=false,bookmarksopen=false,
 breaklinks=false,pdfborder={0 0 0},pdfborderstyle={},backref=false,colorlinks=true]
 {hyperref}



\begin{document}
\preprint{APS/123-QED}
\title{Acoustic properties of metallic glasses at low temperatures -- tunneling
systems and their dephasing\\
\vspace{2mm}
\small{Dedicated to Siegfried Hunklinger on the occasion of his 80$^{\text{th}}$ birthday} }

\author{Arnold Meißner}
\email{meissner.arnold@web.de}
\affiliation{Physikalisches Institut, Karlsruhe Institute of Technology,
Kaiserstraße 12, 76131 Karlsruhe, Germany}
\author{Tim Voigtländer}
\affiliation{Physikalisches Institut, Karlsruhe Institute of Technology,
Kaiserstraße 12, 76131 Karlsruhe, Germany}
\author{Saskia M. Meißner}
\affiliation{Physikalisches Institut, Karlsruhe Institute of Technology,
Kaiserstraße 12, 76131 Karlsruhe, Germany}
\author{Uta Kühn}
\email{U.Kuehn@ifw-dresden.de}
\affiliation{IFW Dresden, Helmholtzstr. 20, 01069 Dresden, Germany}
\author{Susanne Schneider}
\email{sschnei@gwdg.de}
\affiliation{Institut für Geophysik, Abt. Didaktik der Physik, Universität Göttingen,
Friedrich-Hund-Platz 1, 37073 Göttingen, Germany}
\author{Alexander Shnirman}
\affiliation{Institut für Theorie der Kondensierten Materie,
Karlsruhe Institute of Technology, Kaiserstraße 12, 76131 Karlsruhe, Germany}
\affiliation{Institute for Quantum Materials and Technologies,
Karlsruhe Institute of Technology, Hermann-von-Helmholtz-Platz 1, 76344 Eggenstein-Leopoldshafen, Germany}
\author{Georg Weiss}
\email{georg.weiss@kit.edu}
\affiliation{Physikalisches Institut, Karlsruhe Institute of Technology,
Kaiserstraße 12, 76131 Karlsruhe, Germany}

\date{\today}
\begin{abstract}
The low temperature acoustic properties of bulk metallic glasses measured
over a broad range of frequencies rigorously test the predictions
of the standard tunneling model. The strength of these experiments
and their analyses is mainly based on the interaction of the tunneling
states with conduction electrons or quasiparticles in the superconducting
state. A new series of experiments at kHz and GHz frequencies on the
same sample material essentially confirms previous measurements and
their discrepancies with theoretical predictions. These discrepancies
can be lifted by considering more correctly the line widths of the
dominating two-level atomic-tunneling systems. In fact, dephasing caused or mediated by interaction with conduction electrons may lead to particularly
large line widths and destroy the tunneling sytems' two-level character
in the normal conducting state.
\end{abstract}
\pacs{33.15.Ta}
\keywords{Suggested keywords}

\maketitle

The low temperature properties of disordered solids make up a fascinating
thermodynamic combination comprising specific heat, thermal conductivity
and dielectric as well as mechanic susceptibilities\cite{ZellerPohl:TunnelingModell,HunklingerArnold:UltrasonicGlass,Phillips1981:AmorphousSolids,Phillips:TLScollection}.
The standard tunneling model\citep{Phillips1972:TLS,Anderson:StandardTunnelingModel}seems
to account for all these properties in a material-independent, universal
manner establishing its fifty-year-long success despite quantitative
disagreement with some experiments. A particularly remarkable example
is the unexpected crossing of the sound velocities of metallic glasses
measured in the superconducting or in the normal state of the same
sample\cite{Neckel:TunnelingModelIncomplete,Esquinazi:VibratingReedSuperconductingMetGlass}. An experiment is shown in fig. \ref{fig:dv_v-transverse_969M}.
The normal state is enforced by a sufficiently strong magnetic field,
otherwise the sample becomes superconducting below $T_{\text{c}}=\unit[1.39]{K}$.
A first and obviously a too naive - or even incorrect - analysis within
the standard tunneling model predicts the sound velocity of the superconducting
state to mark an upper bound, the sound velocity of the normal state
being always smaller and smoothly merging at some temperatures well
below $T_{\text{c}}$.

In this paper, we discuss all questions related with the acoustic
susceptibility of disordered solids at low temperatures and emphasize
the importance of dephasing of the two-level atomic-tunneling systems
due to their interaction with conduction electrons. We present a series
of new experiments at audio and radio frequencies and demonstrate
that they are very well described within the tunneling model by including
more carefully extreme line widths of the two-level systems in the
normal conducting state of the metallic glass.

\section{Sound velocity in metallic glasses -- stating the problem}

\begin{figure}
\includegraphics{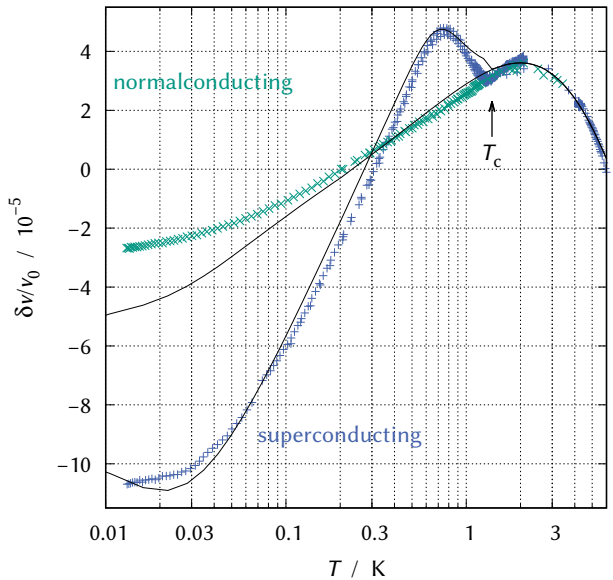}

\caption{\label{fig:dv_v-transverse_969M}Temperature dependence of the transverse
sound velocity of the metallic glass $\text{Zr}_{59}\text{Ti}_{3}\text{Cu}_{20}\text{Ni}_{8}\text{Al}_{10}$
measured at 969\,MHz. The superconducting transition
is at $T_{\text c}=\unit[1.39]{K}$. Data marked \emph{normal conducting} is obtained in a magnetic field of $\unit[4]{T}$. Lines show numerical calculations as explained in the text.}
\end{figure}

The Tunneling Model (TM) provides a microscopic explanation for the low energy excitations
present in virtually all disordered solids. It assumes
that some atoms or groups of atoms in metastable configurations can
occupy two different configurations modelled as particles in double
well potentials, and that at low temperatures the atoms can tunnel
between the configurations. The overlap of the two localized wave
functions generates a two-level system with eigenfunctions across
both the potential wells. Coupling to phonons or strain fields requires
these two-level tunneling systems (TSs) to hold an elastic dipole
(more correctly: quadrupole) moment with different sizes or orientations
for the two configurations. When strain is applied to the material
the tunneling systems may follow this field leading to an increase
of the overall susceptibility of the material, making it 'softer'
than without TSs.

The softening, however, is only effective when a TS is in its ground
state. At higher temperatures, when the population difference vanishes,
its contribution to the susceptibility disappears. At very low temperatures
all relevant TS with large enough energies are in the ground state,
each adding a small contribution to the susceptibility. This so-called
\emph{resonant process} is analogous to a pendulum following in phase
its drive at very low frequency. With increasing temperature both
states of these TSs become equally populated which removes their contributions
to the susceptibility. The material becomes stiffer, the sound velocity
increases. This is visible in fig.~\ref{fig:dv_v-transverse_969M}
where the sound velocity in the superconducting state increases with
temperature from $\unit[50]{mK}$ to $\unit[500]{mK}$. The increase
may persist as long as there are TSs with even higher energies being
thermally excited with rising temperature. 

There is, however, another mechanism by which two-level atomic-tunneling
systems contribute to the acoustic susceptibility: A \emph{relaxation
mechanism}, which becomes important for temperatures where the energy
relaxation rates of the TSs are fast enough so that their occupation
numbers can adjust for the modulation of the energy splitting caused
by the applied strain of the sound field\cite{Jaeckle:UltrasonicAttenuationGlasses}. For relaxation rates much
higher than the sound frequency the occupation of the two levels will
always be close to thermal equilibrium. Due to their continuous relaxation
towards the modulated equilibrium TSs with energies on the order of
the thermal energy contribute to the total susceptibility and reduce
the sound velocity in a certain temperature window. This is particularly
effective for TSs with asymmetric double well potentials since the
lower and higher energy levels are identifiable with the tunneling
particle residing preferentially in either of the two wells.

At this point, possible relaxation processes and the respective temperature
dependence of the resulting relaxation rates ought to be discussed.
Common for all disordered solids and well documented is the relaxation
of TS due to interaction with the phonon bath. At low temperatures
the one-phonon process, i.e. emission or absorption of a single resonant
phonon dominates in adjusting the TSs' occupation numbers\cite{Jaeckle:UltrasonicAttenuationGlasses} whereas
higher order phonon processes become important at elevated temperatures.
In fig.~\ref{fig:dv_v-transverse_969M} (see also fig.~\ref{fig:dv_0.2-1.6GHz})
phonon dominated relaxation is the reason for the sound velocity to
exhibit a maximum around $\unit[2]{K}$ and to further decrease with
rising temperature.

In a normal conducting metallic glass the energy relaxation rates
are massively enhanced because the TSs scatter inelastically conduction
electrons\cite{Golding:RelaxationCondElectrons,Black:TSinMetallicGlass}, which have a high density of states available around the
\noun{Fermi} energy. This additional relaxation channel causes the
sound velocity to be reduced with respect to the always present resonant
process discussed above. The reduction persists to very low temperatures
although the positive temperature coefficient prevails. If the same
sample, i.e. with the same distribution of TSs, became superconducting
below a certain temperature, the electronic relaxation channel would
rapidly vanish and, given that relaxation by phonons is already too
slow, the resonant process alone would determine the temperature dependence
of the sound velocity. It marks an upper bound in the absence of any
relaxation mechanism. Nevertheless, as shown in fig.~\ref{fig:dv_v-transverse_969M}
below $\unit[0.5]{K}$ the sound velocity in the normal conducting sample 
is higher than in the superconducting sample, although relaxation
processes due to inelastic TS-electron scattering are present. The electronic
relaxation becomes also important in the superconducting state when
the density of thermally excited quasiparticles rapidly increases
on approaching $T_{{\rm c}}$. The related contribution to the susceptibility
leads to a maximum of the sound velocity just below $T_{{\rm c}}$
visible in fig.~\ref{fig:dv_v-transverse_969M}.

Summarizing the problem laid out in this section: Despite the increased
susceptibility due to the electronic relaxation process, which in
the normal conducting state persists to lowest temperatures, the material
is at low temperatures stiffer, the sound velocity being higher than
in the superconducting state. As both resonant and relaxation mechanisms
can only increase the susceptibility and thus decrease the sound velocity,
either the number of contributing TSs is reduced in the normal state
or one of the mechanisms is quenched in some way. Renormalization
of the density of states of TS in the normal conducting state\citep{Black:RenormalizationTheoryTLSinMetallicGlass,KaganProkov:Renormalization,BezuglyiZherlitsyn:ElectronRenormalizationBMG}
followed the first approach. However, only a qualitatively better
description of the phenomenon could be achieved. In this paper an
alternative idea is suggested: The contribution of the resonant process
is reduced due to the large linewidth of TSs resulting from the strong
interaction with conduction electrons. While retaining the originally
proposed distribution function (see eq.~\ref{eq:Distribution} further
down) of the Standard Tunneling Model, very satisfying agreement with
experiments at frequencies from kHz to GHz is achieved that way.

\section{\label{sec:level1}The Standard Tunneling Model:\protect \\
a short review}

The standard tunneling model, as introduced by \noun{Phillips}\citep{Phillips1972:TLS}
and \noun{Anderson }et al.\citep{Anderson:StandardTunnelingModel},
and its consequences for the acoustic susceptibility are briefly outlined
here. For most quantities we use the notation as of \noun{Jäckle}\citep{Jaeckle:UltrasonicAttenuationGlasses}.

The tunneling entity is modeled as a particle in a double well potential.
With the WKB-method\footnote{Named after \noun{G. Wenzel}, \noun{H. A. Kramers} and \noun{L. Brillouin}}
the tunneling energy $\varDelta=\hbar\omega_{\text{H}}\exp(-\lambda)$
can be calculated from the tunneling parameter $\lambda=d/2\hbar\cdot\sqrt{2mV}$,
which absorbs the geometric details of the tunneling particle, being
the distance $d$, the mass of the tunneling particle $m$ and the
potential barrier separating the two wells $V$. The asymmetry energy
$\varepsilon$ denotes the energy shift between the ground states
of the individual wells. This TS is described by

\begin{equation}
\mathcal{H}_{0}^{\text{LR}}=\frac{1}{2}\left(\begin{array}{cc}
\varepsilon & -\varDelta\\
-\varDelta & -\varepsilon
\end{array}\right)
\end{equation}
in the localized basis of the particle being in the left state $|L\rangle$
or the right state $|R\rangle$. In the energy basis $\mathcal{H}_{0}^{\text{ge}}=E/2\cdot\sigma_{z}^{\text{ge}}$
with the total energy splitting $E=\sqrt{\varepsilon^{2}+\varDelta^{2}}$.
$\sigma_{\text{z}}^{\text{ge}}$ is the diagonal \noun{Pauli} matrix
in the energy basis $|g\rangle$ and $|e\rangle$ with the tunneling
particle in superposition states of $|L\rangle$ and $|R\rangle$.

External strain $\overleftrightarrow{S}$ applied to the material
changes the asymmetry energy $\delta\varepsilon=\overleftrightarrow{\gamma}\cdot\overleftrightarrow{S}$
of the TSs due to their deformation potential $\overleftrightarrow{\gamma}$.
This coupling is diagonal in the localized basis $\mathcal{H}_{\text{int}}^{\text{LR}}=1/2\,(\overleftrightarrow{\gamma}\cdot\overleftrightarrow{S})\sigma_{\text{z}}^{\text{LR}}$
and yields the interaction Hamiltonian in the energy basis

\begin{equation}
\mathcal{H}_{\text{int}}^{\text{ge}}=\frac{1}{2E}\left(\begin{array}{cc}
\varepsilon & -\varDelta\\
-\varDelta & -\varepsilon
\end{array}\right)\left(\overleftrightarrow{\gamma}\cdot\overleftrightarrow{S}\right)\label{eq:TSinteractionHamilton}
\end{equation}
where for the remaining part of the paper the tensorial notation of
$\overleftrightarrow{\gamma}$ and $\overleftrightarrow{S}$ is dropped
for simplicity. With the transition matrix element 
$M=\gamma\varDelta/2E$ and the longitudinal matrix element $D=\gamma\varepsilon/E$
eq.~(\ref{eq:TSinteractionHamilton}) is rewritten as $\mathcal{H}_{\text{int}}^{\text{ge}}=1/2\cdot(D\sigma_{\text{z}}^{\text{ge}}-2M\sigma_{\text{x}}^{\text{ge}})S$.
The contribution of the TS to the susceptibility $\chi$ of the amorphous
material in a sound field of frequency $\omega$ can be derived from
\noun{Bloch} equations or another linear response theory.

Following e.g. \noun{Hunklinger} and \noun{Arnold}\citep{HunklingerArnold:UltrasonicGlass}
(or \noun{Phillips}\citep{Phillips:TLScollection} for an analogous
discussion in the dielectric case) the response of an ensemble of
$N$ TSs to the transverse interaction in $\mathcal{H_{\text{int}}^{\text{ge}}}$
or the so-called \emph{resonant} \emph{contribution} results in real
and imaginary parts of the susceptibility

\begin{eqnarray}
\chi_{\text{res}}' & = & -4NM^{\,2}\tanh\left(\frac{E}{2k_{\text{B}}T}\right)\label{eq:Chi-Resonant-Real}\\
 & \times & \frac{1}{\hbar}\left(\frac{(\omega-\omega_{0})\tau_{2}^{2}}{1+(\omega-\omega_{0})^{2}\tau_{2}^{2}}-\frac{(\omega+\omega_{0})\tau_{2}^{\,2}}{1+(\omega+\omega_{0})^{2}\tau_{2}^{\,2}}\right)\,,\nonumber 
\end{eqnarray}

\begin{eqnarray}
\chi_{\text{res}}'' & = & 4NM^{\,2}\tanh\left(\frac{E}{2k_{\text{B}}T}\right)\label{eq:Chi-Resonant-Imag}\\
 & \times & \frac{1}{\hbar}\left(\frac{\tau_{2}}{1+(\omega-\omega_{0})^{2}\tau_{2}^{\,2}}-\frac{\tau_{2}}{1+(\omega+\omega_{0})^{2}\tau_{2}^{\,2}}\right)\nonumber 
\end{eqnarray}
where $\omega_{0}=E/\hbar$. The real part $\chi'$ can not be further
simplified by the rotating wave approximation as especially for low
measuring frequencies the contributions at $\omega+\omega_{0}$ and
$\omega-\omega_{0}$ add up and are equal for $\omega\ll\omega_{0}$.
The linewidth $h\tau_{2}^{-1}$ is usually taken to be small. We will
show that this is not always the case and can have a large impact
on the susceptibility of a broad distribution. In contrast to the
real part the loss $\chi_{\text{res}}''$ is dominated by an ensemble
subset with $\omega_{0}\approx\omega$ of the broad distribution of
TSs.

The longitudinal interaction in $\mathcal{H_{\text{int}}^{\text{ge}}}$
gives rise to the \emph{relaxation contribution} to the susceptibility 

\begin{eqnarray}
\chi_{\text{rel}}' & = & -2ND^{\,2}\frac{\partial}{\partial E}\tanh\left(\frac{E}{2k_{\text{B}}T}\right)\frac{1}{1+(\omega\tau_{1})^{2}}\label{eq:Chi_relaxation_real}\\
\chi_{\text{rel}}'' & = & -2ND^{\,2}\frac{\partial}{\partial E}\tanh\left(\frac{E}{2k_{\text{B}}T}\right)\frac{\omega\tau_{1}}{1+(\omega\tau_{1})^{2}}\label{eq:Chi_relaxation_imag}
\end{eqnarray}
where $\partial/\partial E\tanh(E/2k_{\text{B}}T)$ is the change
of the occupation number difference due to variation of the energy
splitting around the given energy $E$. TSs contribute to the longitudinal
susceptibility when their energy relaxation rate $\tau_{1}^{\,-1}$
matches the frequency $\omega$ of the external driving field. In
order to understand the longitudinal contribution knowledge of the
energy relaxation processes is crucial. In turn, the temperature dependence
of the susceptibility gives information on the dominating relaxation
process in a given temperature range.

In metallic glasses TSs relax to the thermal equilibrium through various
processes. Independent of the material being metallic or insulating
the interaction of TSs with the phonon bath is always present and
according to \noun{Jäckle}\citep{Jaeckle:UltrasonicAttenuationGlasses}
the one-phonon process yields the energy relaxation rate

\begin{equation}
\tau_{1,\text{ph}}^{\,-1}=\left(\frac{\gamma_{\text{l}}^{\,2}}{v_{\text{l}}^{\,5}}+2\frac{\gamma_{\text{t}}^{\,2}}{v_{\text{t}}^{\,5}}\right)\left(\frac{\varDelta}{E}\right)^{2}\frac{E^{\,3}}{2\pi\rho\hbar^{4}}\coth\left(\frac{E}{2k_{\text{B}}T}\right)\,.\label{eq:RelaxationsRatePhononen}
\end{equation}
$\rho$ is the density of the material, indices l and t denote the
phonon modes as longitudinal and transverse, respectively. Since $v_{{\rm l}}\approx2v_{{\rm t}}$
and both velocities enter with fifth power in the denominator $\tau_{1,\text{ph}}^{\,-1}$
is by far dominated by transverse phonons. For a given ratio of $\varDelta/E$
thermal TSs ($E\approx k_{\text{B}}T$) have energy relaxation rates
$\propto T^{3}$. At temperatures above $\unit[2]{K}$ multi-phonon
processes lead to energy relaxation rates with stronger temperature
dependencies, modeled for example\citep{Maynard1976:MultiPhonon}
by $\tau_{1,\text{mph}}^{\,-1}=K_{\text{mph}}\cdot T^{5}$.

In a metallic glass the most relevant relaxation process is caused
by the TSs' interaction with conduction electrons which towards lower
temperatures exceeds the phonon-induced rate by many orders of magnitude.
From a microscopic point of view \noun{Black} et al.\citep{Golding:RelaxationCondElectrons,Black:TSinMetallicGlass}
derived the energy relaxation rate

\begin{equation}
\tau_{1,\text{el}}^{\,-1}=W_{\text{el}}^{\,2}\frac{\pi}{4\hbar}\left(\frac{\varDelta}{E}\right)^{2}E\coth\left(\frac{E}{2k_{\text{B}}T}\right)\,.\label{eq:RelaxationRateNormalMetal}
\end{equation}
The constant density of states of electronic excitations around the
Fermi energy yields a linear increase $\propto T$ of the rate for
thermal TS.

In the superconducting state electrons form Cooper pairs and well
below $T_{{\rm c}}$ only the phonon induced rate remains. However,
with rising temperature the interaction of TSs with thermally excited
quasiparticles leads to a rapidly increasing rate approaching the
normal conducting rate\citep{Black:SuperconductingMetallicGlassTLS,Black:TSinMetallicGlass}
at $T_{\text{c}}$
\begin{equation}
\tau_{1,\text{qp}}^{\,-1}=W_{\text{el}}^{\,2}\frac{\pi}{2\hbar}\negthickspace\left(\frac{\varDelta}{E}\right)^{\negthickspace2}\negthickspace E\coth\negmedspace\left(\frac{E}{2k_{\text{B}}T}\right)\negthickspace\frac{1}{1+\exp(\frac{\Delta_{\text{BCS}}(T)}{k_{\text{B}}T})}\negthickspace\negthickspace\label{eq:RelaxationRateSuperconductor}
\end{equation}
This relation holds strictly only for $E\ll\Delta_{\text{BCS}}$.
Closely below $T_{{\rm c}}$ pair breaking effects are possible \citep{Black:SuperconductingMetallicGlassTLS}
and may result in corrections of our calculations.

The rates $\tau_{1}^{\,-1}(T)$ of all three processes (\ref{eq:RelaxationsRatePhononen}),
(\ref{eq:RelaxationRateNormalMetal}), and (\ref{eq:RelaxationRateSuperconductor})
are shown as solid lines in fig.~\ref{fig:Energy-relaxation-rate}
further down together with experimental data extracted from the position
of the maxima in the sound velocity, which marks the temperature where
$\tau_{1}^{\,-1}\approx\omega$, as elucidated below. The TS-electron
coupling strength $W_{\text{el}}$ is determined from the $\delta v/v$
maximum in the superconducting sample via eq.~(\ref{eq:RelaxationRateSuperconductor}).
Inserting this value for the electron-induced relaxation rate in the
normal conducting state, eq.~(\ref{eq:RelaxationRateNormalMetal}),
one finds rates exceeding the energy of the thermal TSs as visualized
in fig.~\ref{fig:Energy-relaxation-rate}.

\begin{figure}
\includegraphics{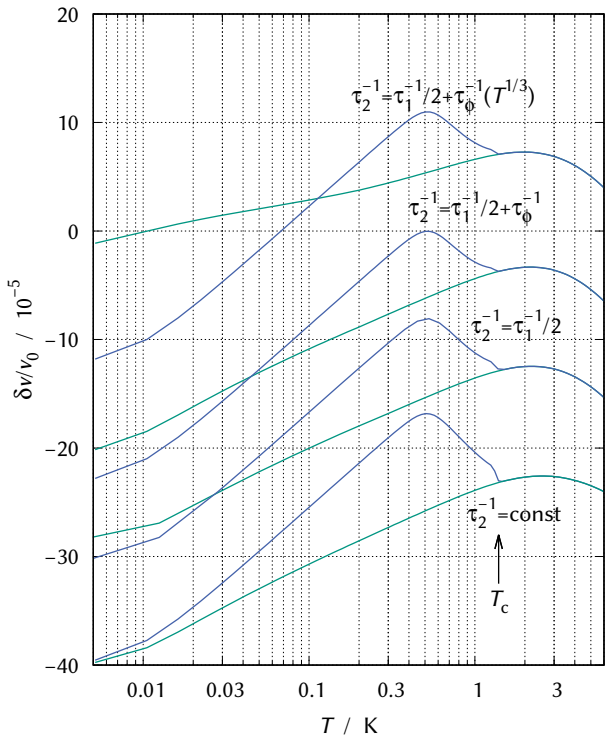}

\caption{\label{fig:DephasingModel}Temperature dependence of the sound velocity
of a metallic glass as calculated with the standard tunneling model
for a measuring frequency of $\unit[200]{MHz}$. Bottom curves comprise
energy relaxation only by phonons (red), by quasiparticles (blue),
and electrons (green). Constant linewidth $h\tau_{2}^{\,-1}\ll E$
is assumed. Sound velocities of normal- and superconducting states
merge at low temperatures. Including the lifetime limited linewidth
$\tau_{2}^{\,-1}=\tau_{1}^{\,-1}/2$ with plausible and temperature
dependent $\tau_{1}^{\,-1}$ rates already produces a crossing of
$\delta v/v$ of the normal state and the superconducting state (curves
next to bottom). Addition of a dephasing process $\tau_{\phi}^{\,-1}\propto T$
shifts this crossing to higher temperatures. Finally, a modified dephasing
model $\tau_{\phi}^{\,-1}\propto f(T)$ (see text) reproduces the
experimental data (see figs \ref{fig:dv_v-transverse_969M} and \ref{fig:dv_0.2-1.6GHz})
very well (topmost curves).}
\end{figure}

Obviously, the linewidth $h\tau_{2}^{\,-1}$ of these dominating,
thermal TSs can not be assumed to be small and constant and at least\citep{WangsnessBloch:NuclearInduction,Redfield:RelaxationNuclearSpin}
$\tau_{2}^{\,-1}=\tau_{1}^{\,-1}/2$ has to be introduced when analyzing
experiments measuring $\chi_{\text{res}}$. The contribution to $\chi_{\text{res}}$
is reduced or even lost, when the phase of a TS is lost within one
oscillation of the external field. Including this minimal requirement
in calculating expected temperature dependencies of $\delta v/v$
on the basis of eqs.~(\ref{eq:Chi-Resonant-Real}) and (\ref{eq:Chi_relaxation_real}),
however replacing $N$ by an integration over the distribution functions
$P(E,\varDelta)$ as proposed in the Standard Tunneling Model (eq.~\ref{eq:Distribution}
further down) one finds already an answer to the major question stated
above: At lowest temperatures the sound velocity in the normal state
is predicted to be higher than in the superconducting state (see fig.~\ref{fig:DephasingModel}).
Obviously, the short life times $\tau_{1}$ already broaden the line
widths of relevant TSs sufficiently to reduce their contributions
to $\chi_{\text{res}}'$, eq.~(\ref{eq:Chi-Resonant-Real}), leaving
the material harder although both resonant and relaxation contributions
are present.

It is in fact sensible to accept explicitly a dephasing process due
to the TS-electron interaction as already suggested by \noun{Black}\citep{Black:TSinMetallicGlass}
when briefly discussing the longitudinal coupling between TS and excitations
of the electron bath. A more general formulation of dephasing of two-level
systems\citep{Shnirman:DissipationQubits} considers explicitly the importance of the spectral density
of the bath in the low energy limit $E\ll k_{{\rm B}}T$.
This leads to a dephasing rate

\begin{equation}
\tau_{\phi,\text{el}}^{\,-1}=W_{\text{el}}^{\,2}\frac{\pi}{2\hbar}\left(\frac{\varepsilon}{E}\right)^{2}k_{\text{B}}T\label{eq:DephasingRateNormalMetal}
\end{equation}
for the spectral density of the electronic bath in the normal conducting
state, which may be modified as

\begin{equation}
\tau_{\phi,\text{qp}}^{\,-1}=W_{\text{el}}^{\,2}\frac{\pi}{\hbar}\left(\frac{\varepsilon}{E}\right)^{2}k_{\text{B}}T\frac{1}{1+\exp(\frac{\Delta_{\text{BCS}}(T)}{k_{\text{B}}T})}\label{eq:DephasingRateSuperconductor}
\end{equation}
for the superconducting state. The magnitude of $\tau_{\phi}^{\,-1}$
is determined by the same coupling constant $W_{\text{el}}$ as $\tau_{1}^{\,-1}$,
and $\tau_{\phi}^{\,-1}$ has the same temperature dependence as $\tau_{1}^{\,-1}$
of thermal TSs with $E\approx k_{\text{B}}T$. But in contrast to
$\tau_{1}^{\,-1}$ dephasing $\tau_{\phi}^{\,-1}$ due to this process
vanishes for symmetric\footnote{A $(\varepsilon/E)^{2}$ dependence of $\tau_{\phi}^{\,-1}$, originating
from the TSs' dipolar interactions, has been observed in dielectric
echo experiments on individual TSs in the oxide films of Josephson
junction qubits \citep{Lisenfeld:DecoherenceSpectroscopySREP}.} TSs with $\varDelta=E$. However, slightly asymmetric TSs are sufficiently affected as well, and their
contribution to $\chi_{\text{res}}'$ is also reduced. Calculation of $\delta v(T)/v$
with\citep{WangsnessBloch:NuclearInduction,Redfield:RelaxationNuclearSpin} $\tau_{2}^{\,-1}=\tau_{1}^{\,-1}/2+\tau_{\phi}^{\,-1}$ produces
a small but noticeable shift to higher temperature of the crossing
between normal and superconducting states (see fig.~\ref{fig:DephasingModel}).

The calculated influence of dephasing can be amplified by replacing
the linear $T$ dependence in eqs.~\ref{eq:DephasingRateNormalMetal}
and \ref{eq:DephasingRateSuperconductor} by $f(T)=T\cdot(T/T_{0})^{\alpha-1}$.
With $0<\alpha<1$ dephasing $\tau_{\phi}^{\,-1}$ varies weaker than
linearly with $T$ and is effectively enhanced towards lower temperatures.
Combination of large $\tau_{1}^{\,-1}$ and strong dephasing $\tau_{\phi}^{\,-1}$
weakens the contribution of symmetric and slightly asymmetric TSs
to $\chi_{\text{res}}'$ and the calculated $\delta v(T)/v$ exhibits
a further reduced temperature variation in the normal conducting metal
shifting the intersection with the superconducting state closer to
$T_{\text{c}}$ (fig.~\ref{fig:DephasingModel} topmost curves).
To this point the suggested temperature dependence $f(T)$ of the
electron-mediated dephasing is purely phenomenological. We will come
back to this later.

Calculation of the response to an applied sound field of an ensemble
of atomic-tunneling systems in a disordered solid requires both, resonant
and relaxation contributions to the susceptibility to be integrated
over the distribution of parameters of all TSs. The standard tunneling
model\citep{Anderson:StandardTunnelingModel,Phillips1972:TLS} suggests
a flat distribution in asymmetry energy $\varepsilon$ and tunneling
parameter $\lambda$ yielding the distribution\citep{Phillips:TLScollection}

\begin{equation}
P(E,\varDelta)\,\text{d}E\,\text{d}\varDelta=\frac{P_{0}E}{\varDelta\sqrt{E^{\,2}-\varDelta^{2}}}\,\text{d}E\,\text{d}\varDelta\label{eq:Distribution}
\end{equation}
of TSs with energy splitting $E$ and tunneling energy $\varDelta$,
where $P_{0}$ is constant. The susceptibility of a macroscopic sample
is given by
\begin{equation}
\langle\chi\rangle=\intop_{0}^{E_{\text{max}}}\text{d}E\intop_{\varDelta_{\text{min}}}^{E}\text{d}\varDelta\,(\chi_{\text{res}}+\chi_{\text{rel}})\,P(E,\varDelta)\,.
\end{equation}
The real part of $\langle\chi\rangle$ yields the change of sound
velocity and the imaginary part the absorption of sound waves, as
quantified in the next paragraph.

We briefly discuss the role of the integration limits, because in
contrast to the known analytic solutions\citep{Phillips:TLScollection}
they don't appear in our final results. The integration limit $E_{\text{max}}$
vanishes when with eq.~\ref{eq:Chi-Resonant-Real} the line width
is adequately taken into account. At high temperatures the energy
relaxation rate caused by interaction with phonons leads to a linewidth
$h\tau_{2}^{-1}$ increasing faster than linearly with energy $E$.
TSs with $h\tau_{2}^{-1}\ge E$ don't contribute to $\chi'_{\text{res}}$
and this limits the increase of the sound velocity above a certain
temperature thus $E_{\text{max}} \rightarrow \infty$ can be taken. Already for the one-phonon process the energy-relaxation
limited linewidth $h\tau_{2}^{-1}=h\tau_{1}^{-1}/2$ scales $\propto\,E^{\,3}$,
for multi-phonon processes the linewidth increases even stronger with
exponents from $E^{\,5}$ to $E^{\,7}$. The upper bound for TSs energies
contributing to $\chi_{\text{res}}'$ is of the order of $\unit[10]{K}$
and almost independent of the prefactors in eq.~\ref{eq:RelaxationsRatePhononen}.
This differes from introducing a high energy cutoff for the distribution
of TSs. Although unimportant for $\chi'_{\text{\text{res}}}$ high
energy TSs are still present and are responsible for the relaxation
contributions to the susceptibility at higher temperatures.

The lower cutoff for the tunneling energy $\varDelta_{\text{min}}$
is required in analytic calculations to avoid the divergence of the
integral for $\varDelta\rightarrow0$. For the transverse susceptibility
this is canceled by $\varDelta$ in the matrix element $M$ representing
the overlap of the localized wave functions and thus defining the
two-level character of a TS with given energy $E$. The longitudinal
susceptibility vanishes, when the temperature dependence of the relaxation
rate is taken into account: Asymmetric TS with $\varDelta\rightarrow0$
become too slow to relax within the time scale of the experiment set
by the measuring frequency and they no longer contribute to $\langle\chi\rangle$.

\section{Experimental Methods}

The TSs' contribution to the real part of the generalized susceptibility
$\langle\chi\rangle'$ is related to the variation of the sound velocity
$\delta v = v(T)-v_0$ normalized to the velocity at a reference temperature
$v_{0}=v(T_{0})$ as\citep{HunklingerArnold:UltrasonicGlass,Phillips:TLScollection}
\begin{equation}
\frac{\delta v}{v}=-\frac{1}{2}\frac{1}{\rho v_0^{\,2}}\langle\chi\rangle'\,.
\end{equation}

The imaginary part $\langle\chi\rangle''$ is related to the loss
and ultrasonic measurements yield the attenuation $\alpha$ per unit
length. Normalized to the energy loss per wavelength results in the loss
tangent\citep{HunklingerArnold:UltrasonicGlass,Phillips:TLScollection}
\begin{equation}
\tan \delta=\frac{\alpha v_{0}}{\omega}=\frac{1}{\rho v_{0}^{\,2}}\langle\chi\rangle''\,.
\end{equation}
In experiments with vibrating reeds or other acoustic resonators the corresponding
quantity is the inverse quality factor $Q^{\,-1}=\tan \delta$ defined
as the energy loss per cycle divided by the total energy stored in
the resonator.

The samples measured in this work are made from one rod (diameter
$\approx\unit[3.5]{mm}$) of the bulk metallic glass $\text{Zr}_{59}\text{Ti}_{3}\text{Cu}_{20}\text{Ni}_{8}\text{Al}_{10}$
fabricated at \noun{IFW Dresden}. Details on the fabrication of
the metallic glass sample and its properties can be found elsewhere\citep{Kuehn:ZrTiCuNiAl}.
At low frequency ($\unit[1.1]{kHz}$) quality factor and variation
of the sound velocity -- determined by tracking the resonance frequency
-- are measured by a capacitively driven vibrating reed\citep{Berry:IBMVibratingReed,Classen:LowFreqAcousticDielectricGlass,Luethi:Physical_Acoustics_in_the_solid_State}
machined as a stripe (rough size $\unit[0.3]{mm}\times\unit[2]{mm}\times\unit[12]{mm}$)\citep{Doettling:VibratingReed}
from one part of the rod. The reed is clamped to a copper sample holder
to provide thermalization in a dilution cryostat. Details of the setup
used in this work are described elsewhere\citep{Seiler:MetGlasSound}.
To investigate the normal conducting state of the material the reed
is exposed to a magnetic field of $\unit[4]{T}$, sufficient to reliably
suppress superconductivity of the metallic glass. In order to minimize
possible eddy currents and other disturbing effects the sample is
mounted such that the direction of its oscillation lies parallel to
the magnetic field vector. 

For the high frequency experiments cylinders were cut from the same
sample rod. Their two end faces were lapped plain and parallel\footnote{Lapping was done by \noun{Stähli-Läpptechnik}, Weil im Schönbuch,
Germany}. RF sputtered zinc oxide transducers of
thicknesses between $\unit[3.6]{\mu m}$ and $\unit[1.2]{\mu m}$
were employed to generate and detect ultrasonic pulses between
$\unit[200]{MHz}$ and $\unit[1.7]{GHz}$ with a phase sensitive homodyne
technique\citep{Seiler:MetGlasSound}. Sample cylinders
were mounted in a copper clamp and the transducers were contacted
by spring loaded probe pins. Longitudinal polarization at $\unit[1.44]{GHz}$
was measured in reflection with only one transducer. All measurements
with transverse polarization were carried out in transmission with a transducer at either
end. This setup allows for the use of cryogenic amplifiers and reduced
noise.

\section{Results and Discussion}

\begin{figure}
\includegraphics{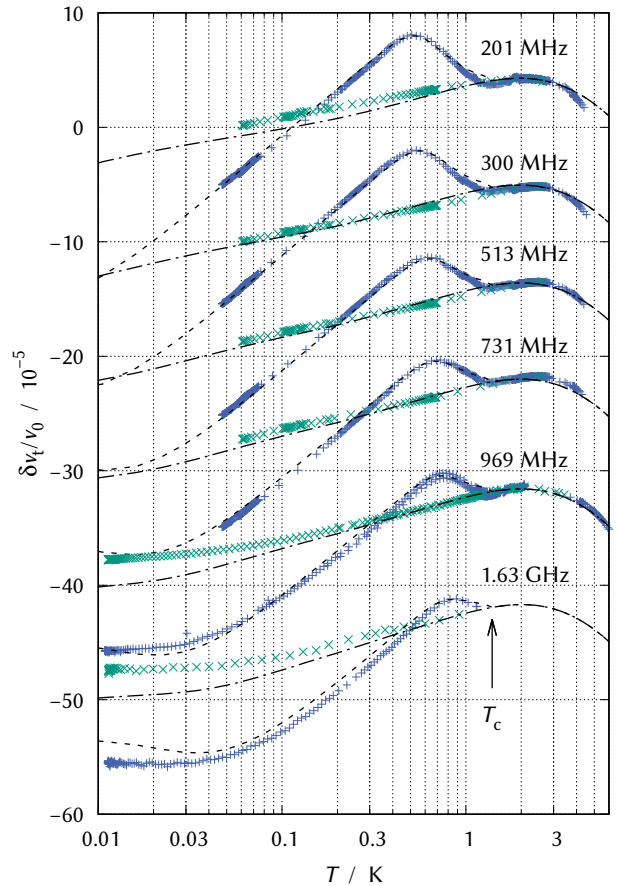}

\caption{\label{fig:dv_0.2-1.6GHz}Relative temperature variation of the transverse
sound velocity of glassy $\text{Zr}_{59}\text{Ti}_{3}\text{Cu}_{20}\text{Ni}_{8}\text{Al}_{10}$
measured at frequencies between $\unit[200]{MHz}$ and $\unit[1.63]{GHz}$.
Blue crosses are data without magnetic field where the sample is superconducting
below $T_{\rm c}=\unit[1.39]{K}$. A magnetic field of $\unit[4]{T}$
keeps the sample normal conducting at all temperatures (green $\times$).
Dashed lines result from numerical calculations using always the same
parameters as detailed in the text. Sets of data and curves for different
frequencies are shifted for clarity.}
\end{figure}

Measurements of the temperature dependence $\delta v/v$ of the sound
velocity of transverse ultrasound is shown in fig.~\ref{fig:dv_0.2-1.6GHz}.
At these frequencies in the GHz range all features caused by TSs with
a broad distribution of parameters as discussed in the previous sections
are clearly observed. Although both resonant and relaxation mechanisms
contribute to the acoustic susceptibility the sound velocity at lowest
temperature is higher in the normal state than in the superconducting
state. The onset of the relaxation process by quasiparticles in the
superconducting state is testified by the pronounced maxima between
0.6 and 0.9\,K depending on frequency. Maxima at $\unit[2]{K}$
demonstrate the frequency independent transition from electron dominated
to phonon dominated relaxation which has been found and described
in earlier experiments\citep{Raychaudhuri:ElasticPropGlasses,Esquinazi:VibratingReedSuperconductingMetGlass}.
Over some temperature ranges the sound velocity varies logarithmically
with temperature. The 'slopes' of the $\log T$ dependencies are a
measure of the TSs' density of states as will be explicated below.

\begin{figure}
\includegraphics{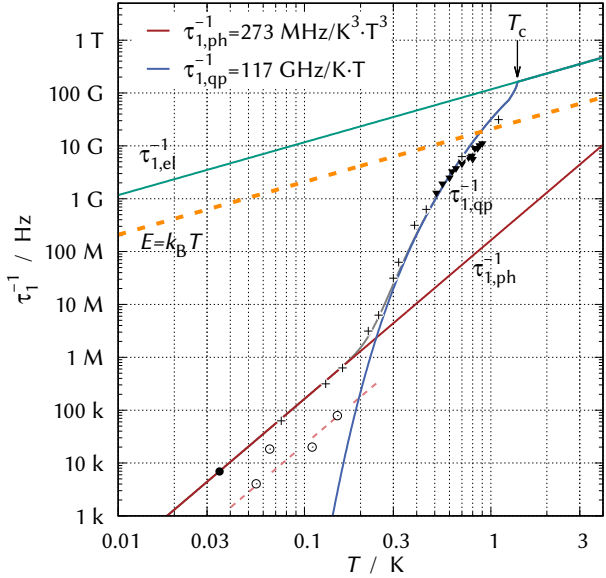}

\caption{\label{fig:Energy-relaxation-rate}Energy relaxation rate of TS in
the metallic glass as extracted from the onset of the relaxation process
related to respective maxima of $\delta v/v$. The energy relaxation
rates $\tau_{1}^{-1}$ of the fastest, symmetric thermal TS due to
interaction with phonons (red), thermally activated quasiparticles
(blue), and electrons in the normal conducting state (green) are adjusted
to the data. The dashed (orange) line indicates the energy splitting
of thermal TS with $E\approx k_{\text{B}}T$. Low frequency data are
determined by vibrating reed experiments, high frequency by pulsed
ultrasound. Full symbols denote data of samples machined from a bulk
metallic glass rod of $\text{Zr}_{59}\text{Ti}_{3}\text{Cu}_{20}\text{Ni}_{8}\text{Al}_{10}$,
open symbols result from of a splat quenched sample
of Zr$_{46.8}$Ti$_{8.2}$Cu$_{7.5}$Ni$_{10}$Be$_{27.5}$. For crosses see text.} 
\end{figure}

In a first step of analyzing the data the frequency dependence of
the $\delta v(T)/v$ maximum in the superconducting state of the sample
is considered. At temperatures below the maximum, the contribution
of the longitudinal susceptibility (\emph{relaxation process}) is
negligible and $\delta v(T)/v$ is solely determined by the transverse
susceptibility (\emph{resonant process}). The contribution of the
longitudinal susceptibility becomes important when the TSs' energy
relaxation rates become larger than the frequency of the applied external field, $\tau_{1}^{-1}\gtrsim\omega$, yielding a maximum in
$\delta v(T)/v$ when the contributions per temperature interval of
the two mechanisms cancel each other. Thus we may plot the frequency
of the ultrasound versus the temperature of the maximum in fig.~\ref{fig:Energy-relaxation-rate}
to extract the temperature dependence of the energy relaxation rate
of thermal TSs. The resulting rate varies rapidly with temperature
between $\unit[0.5]{}$ and $\unit[0.9]{K}$ and is clearly identified with the
relaxation rate of eq.~(\ref{eq:RelaxationRateSuperconductor}) caused
by the TSs' interaction with quasiparticles and plotted as blue line in fig.~\ref{fig:Energy-relaxation-rate}. 
In a better quantitative analysis respective maxima in $\delta v(T)/v$ are extracted from numerical calculations including all contributions. Plotted as black crosses in fig.~\ref{fig:Energy-relaxation-rate}, these calculated maxima clearly confirm this analysis and yield an interaction strength $K_{\text{el}}=W_{\text{el}}^2 k_{\text{B}} \pi/2\hbar=\unit[117]{GHz/K}$ in the superconducting state.

\begin{figure}
\includegraphics{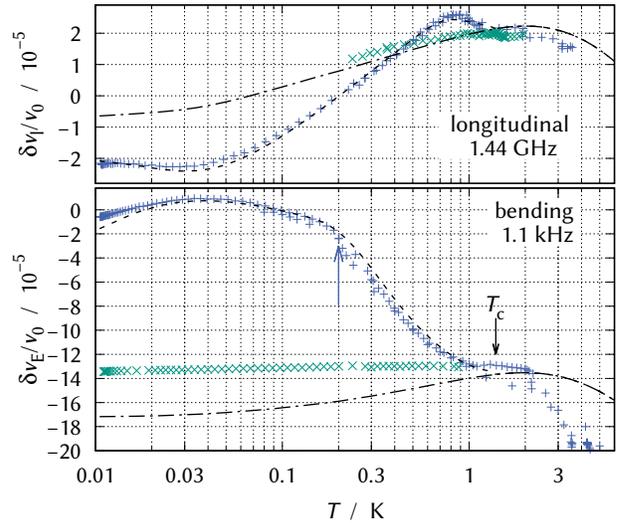}

\caption{\label{fig:dv_reed_vt_vl}Temperature dependence of the longitudinal
sound velocity at $\unit[1.44]{GHz}$ (upper panel) and of the Young's
modulus velocity measured by a vibrating reed\cite{Adler:VibReedCJP2011} at $\unit[1.1]{kHz}$
(lower panel).
Arrow indicates the transition from phonon to quasiparticle
dominated relaxation. Dashed lines show numerical calculations.}
\end{figure}

As already claimed in the previous section, transfer of the same value $K_{\text{el}}=\unit[117]{GHz/K}$ to the normal conducting state results in relaxation
rates eq.~(\ref{eq:RelaxationRateNormalMetal}) of the dominating thermal TSs which considerably exceed their energy splitting
divided by $h$ as depicted by the green line in fig.~\ref{fig:Energy-relaxation-rate}.
Thus their contribution to the transverse susceptibility is reduced,
the sound velocity remains at higher values. Continuing this line
of argument it seems plausible to introduce a regular dephasing based
on TS-electron interaction with respective rates as given in eqs.~(\ref{eq:DephasingRateNormalMetal})
and (\ref{eq:DephasingRateSuperconductor}). In a last step the linear
temperature dependence of these rates is replaced by $f(T)=T\cdot(T/T_{0})^{\alpha-1}$
and it turns out that with $\alpha=1/3$ the overall best agreement
of numerical calculations with experiment is achieved. Apart from
taking into account the polarization of the deformation being transverse
or longitudinal all relevant parameters concerning distribution and
dynamics of the TSs are the same for the calculated graphs of figs.~\ref{fig:dv_0.2-1.6GHz} and \ref{fig:dv_reed_vt_vl}.

Although it is evident that dephasing plays the key role in understanding the dynamics and susceptibility of TSs in metallic glasses our analysis has some deficiencies. In the normal state both resonant and relaxation contributions make up the total temperature dependence of $\delta v/v$. Experimentally $\delta v/v$ is a straight $\log T$ which is not perfectly reproduced by our caluculation. This is noticable with the 201\,MHz data in fig.~\ref{fig:dv_0.2-1.6GHz}, however most clearly visible for the 1.1\,kHz data in fig.~\ref{fig:dv_reed_vt_vl}. We'll come back to this point. 

In high frequency experiments a minimum of the sound velocity is expected
at a temperature where ${\rm h}\nu\approx2.2\,k_{\rm B}T$. In a small temperature range with ${\rm h}\nu\gtrsim E\approx k_{\rm B}T$ a negative contribution to the susceptibility $\chi_{\text{res}}'$ leads to an inverted temperature dependence before this mechanism 
becomes ineffective\cite{HunklingerArnold:UltrasonicGlass} at even lower temperatures where $k_{\rm B}T\approx E\ll{\rm h}\nu$.
In our measurements (see fig.~\ref{fig:dv_0.2-1.6GHz}) these minima
don't appear as pronounced as predicted. The reasons might be that
particularly in the superconducting state the sample was not cooled
sufficiently at lowest temperatures and that in the normal state this
effect is additionally smeared out by the relaxation contribution.
For high frequencies, the condition $\omega\tau_1\gtrsim1$ may be reached after all at lowest temperatures in the normal conducting state,
leading to a flattening of the relaxation contribution to $\delta v(T)/v$. 

Figure \ref{fig:dv_reed_vt_vl} shows two more measurements of the
sound velocity change versus temperature, one with longitudinal ultrasound
at $\unit[1.44]{GHz}$ and a vibrating reed experiment at $\unit[1.1]{kHz}$
probing essentially longitudinal deformation as well. In the superconducting state the low frequency experiment exhibits a maximum around $\unit[35]{mK}$
which is identified as onset of the relaxation contribution. Since
$T\ll T_{\text{c}}$ the quasiparticle induced rate is certainly negligible.
The one-phonon process dominates the energy relaxation of TSs and symmetric TSs have a relaxation rate 
$\tau_{1}^{\,-1}=1.67\,\omega$ at the maximum. We may thus add another
important data point to fig.~\ref{fig:Energy-relaxation-rate} (full
circle) and with eq.~(\ref{eq:RelaxationsRatePhononen}) we may extract
the deformation potential of the TS with respect to transverse phonon
modes from this experiment. We find $\tau_{1}^{-1}=\unit[273]{MHz/K^{3}}\cdot T^{3}$
for the fastest TSs corresponding to a deformation potential $\gamma_{\text{t}}=\unit[0.49]{eV}$.
Further independent measurements of another vibrating reed sample
fabricated from splat-quenched Zr$_{46.8}$Ti$_{8.2}$Cu$_{7.5}$Ni$_{10}$Be$_{27.5}$ (Vitreloy 4)\cite{SchneiderSun:OxidationJMR1996} allow to
determine relaxation rates between $\unit[0.6]{kHz}$ and $\unit[12]{kHz}$
in an analogue way (open circles in fig.~\ref{fig:Energy-relaxation-rate}).
The frequency dependent shift supports the $T^{3}$ increase of the
relaxation rate as predicted by eq.~(\ref{eq:RelaxationsRatePhononen}),
however with a smaller $\gamma_{\text{t}}$.

In fig.~\ref{fig:Energy-relaxation-rate}, the relaxation rate exhibits at $\unit[200]{mK}$ a transition from phonon-dominated energy relaxation of the TSs to relaxation dominated by the interaction with quasiparticles. This is clearly reproduded in the $\delta v(T)/v$ measurement at $\unit[1.1]{kHz}$ (see fig.~\ref{fig:dv_reed_vt_vl}) as a distinct change of the temperature dependence marked by an arrow.

Quantitative inspection of the sound velocity measurements (figs.\,\ref{fig:dv_0.2-1.6GHz} and \ref{fig:dv_reed_vt_vl}) in the superconducting
state allows to extract the density of states of TSs. Analytical
integration of the susceptibility's resonant contribution $\chi_{\text{res}}^{\,-1}$
with $P(E,\varDelta)$ yields a logarithmic temperature dependence
of $\delta v/v$ for $T>\text{h}\nu/2.2k_{\text{B}}$ in agreement
with experiments as long as the contribution of relaxation is
negligible \citep{HunklingerArnold:UltrasonicGlass}

\begin{equation}
\frac{\delta v(T)}{v}|_{{\rm res}}=\frac{P_{0}\gamma_{\text{l,t}}^{2}}{\rho v_{\text{l,t}}^{2}}\cdot\ln\frac{T}{T_{0}}\label{eq:logSlope}
\end{equation}

Numerical calculations reproduce this behavior most clearly for the
measuring frequencies between $201$ and $\unit[513]{MHz}$. The magnitude
of the logarithmic slope $C_{\text{l,t}}=P_{0}\gamma_{\text{l,t}}^{2}/\rho v_{\text{l,t}}^{2}$
contains the mass density of the material\citep{Kuehn:ZrTiCuNiAl}
$\rho=\unit[6592]{kg/m^{3}}$ and the transverse and longitudinal
sound velocities $v_{\text{t}}=\unit[2182]{m/s}$ and $v_{\text{l}}=\unit[4741]{m/s}$
which were determined independently. From the transverse ultrasound
data in the superconducting state of fig.~\ref{fig:dv_0.2-1.6GHz}
we extract an average and frequency independent $\ln(T/T_0)$ slope of
$C_{\text{t}}=5.9\cdot10^{-5}$ and knowing $\gamma_{\text{t}}$ from
the vibrating reed measurement we find $P_{0}=C_{\text{t}}\rho v_{\text{t}}^{2}/\gamma_{\text{t}}^{2}=\unit[3.0\cdot10^{44}]{J^{-1}m^{-3}}$. Analog evaluation of fig.~\ref{fig:dv_reed_vt_vl}
yields $C_{\text{l}}=2.07\cdot10^{-5}$ for longitudinal ultrasound.
The vibrating reed experiment allows to extract a value $C_{\text{E}}=4.17\cdot10^{-5}$
(related to the Young's modulus deformation field) from the total $\delta v(T)/v$
decrease between $\unit[40]{mK}$ and $T_{{\rm c}}$ dominated by the
relaxation contribution to $\chi'$, and also but minor importantly by a decreasing resonant contribution. $C$ and $P_{0}$ values resemble
those of other dielectric and metallic glasses\cite{UniversalAcousticProps,Doussineau:FlourideGlass,Classen:LowFreqAcousticDielectricGlass}.

Closer inspection of $\delta v(T)/v$ in the superconducting state
reveals that the $\ln(T/T_0)$ behavior has a slight upward bend. Quantitatively for the 969\,MHz measurements (see also supplemental material \citep{Suppl}), and quite similar for other frequencies, the the slope increases from $C_{\text{t}}=5.2\cdot10^{-5}$ in the temperature
range between $\unit[100]{mK}$ and $\unit[200]{mK}$ to about $C_{\text{t}}=6.2\cdot10^{-5}$
between $\unit[300]{mK}$ and $\unit[500]{mK}$, still well below
the maximum (see fig.\,1 in supplemental material\citep{Suppl}). This suggests an equivalent increase of less than 20\% of
the TSs' density of states $P_{0}$ in the corresponding energy range
$E\approx k_{\text{B}}T$. A reduced density of states at lower energies
might be explained by models invoking a so-called \emph{dipole gap}
arising from dipole-dipole interaction between TSs\citep{Burin:CollectiveExcitationGlass,Burin:DipoleGapRelaxation},
which becomes especially important at very low temperatures and frequencies. However, thermal conductivity measurements on another Zr based superconducting
bulk metallic glass do not support the idea of a reduced density
of states of TSs towards lower energies\citep{Rothfuss:ThermalPropBMG,Rothfuss:ThermCondBMG} -- there seems to be no clear answer yet for this class of disordered matter.

In contrast to the slight temperature dependence of the slope in the
superconducting state, a clear $\log T$ temperature dependence of
$\delta v/v$ is always observed in the material's normal conducting state. It is further observed that the slope depends on frequency. It decreases by 15\%
when lowering the measuring frequency from $\unit[731]{MHz}$ to $\unit[201]{MHz}$.
The extremely small, though still $\log T$ temperature dependence
of $\delta v/v$ at $\unit[1.1]{kHz}$ is remarkable (fig.~\ref{fig:dv_reed_vt_vl})
and requires a delicate balance of resonant and relaxation contributions.
Previous vibrating reed measurements on glassy PdZr and CuZr alloys\citep{Esquinazi:VibratingReedSuperconductingMetGlass}
show an overall quite similar behavior, however, even a negative slope
of $\delta v(T)/v$ is observed in a temperature range where a positive
slope is expected. Quantitatively, this frequency dependence is not understood. It seems to be plausible, however, that short dephasing
times disturbing the TSs' coherence and eventually leading to localization
are more sensitively detected the lower the measuring frequency is. In extreme cases the still present relaxation contribution may dominate and lead to a negative temperature dependence of total $\delta v/v$.

\begin{figure}
\includegraphics{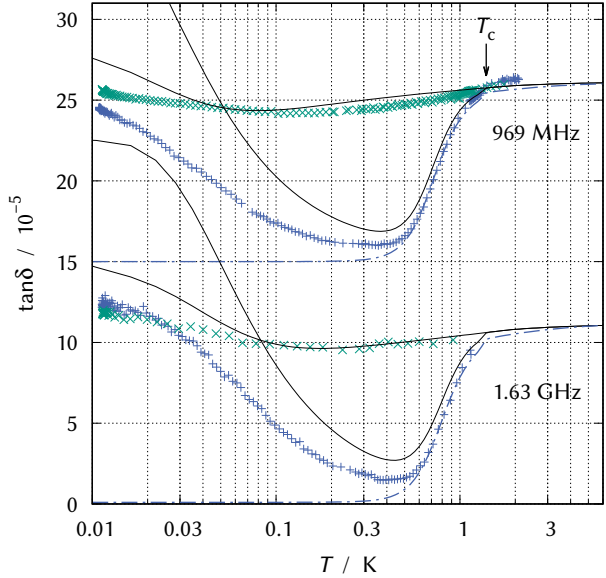}

\caption{\label{fig:loss-tangent_969M+1G6}Temperature dependence of the attenuation $\tan \delta$ of transverse ultrasound in the GHz frequency range. Values are determined by the squared ratio of the amplitudes of the first transmitted ultrasound pulse and a fixed reference signal.  
Solid lines are calculated for weak excitation using values for $W_{\text{el}}$ and $\gamma_{\text{t}}$ derived from the temperature dependence of the sound velocity. For $P_0$ values see text. The acoustic intensity for the experiment was not at the weak excitation level. Dash-dotted lines show calculations of only the relaxation contribution of the superconduction sample. Experimental data of both frequencies is vertically shifted to adjust for the zero line, $\unit[969]{MHz}$ data are shifted by another $15\cdot10^{-5}$.}
\end{figure}

In this last section, the attenuation of transverse ultrasound is discussed. As for the temperature dependence of the sound velocity, the attenuation due to TSs comprises both resonant and relaxation contributions and numerical calculation requires integration of the corresponding imaginary parts of the susceptibility, $\chi_{\text res}''$ and $\chi_{\text rel}''$.

The resonant absorption (transverse interaction) selects from the broad distribution those TSs for which $E=\hbar\omega$ and its temperature dependence reflects the respective occupation number difference $\propto\tanh(E/2k_{\text{B}}T)$. The temperature dependence of the relaxation absorption (longitudinal interaction) -- in particular its increase when coming from low temperatures -- is determined by the dominating relaxation process, caused by interaction with phonons, electrons, or quasiparticles and by the measuring frequency. In any case, $\tan\delta$ arrives at a frequency and temperature independent plateau\cite{Jaeckle:UltrasonicAttenuationGlasses,Black:SuperconductingMetallicGlassTLS} when the energy relaxation of the fastest TSs attains the condition $\omega\tau_1\approx1$.

Figure \ref{fig:loss-tangent_969M+1G6} shows measurements at 969\,MHz and 1.63\,GHz together with respective theoretical curves which are calculated with parameters $\gamma_{\text{t}}$ and $W_{\text{el}}$ as extracted from the $\delta v/v$ measurements. $P_0$ was magnified by 20\%. Same parameters are used for all frequencies (more frequencies shown in fig.\,2 in supplemental material\citep{Suppl}). Constant values have been subtracted from the complete experimental data sets of the two frequencies to adjust for the zero line given by the calculation. Resonant and relaxation contributions are easily identified, particularly in the superconducting state. There, the measured resonant absorption exhibits the correct temperature dependence, however, is smaller than predicted by roughly a factor two. This discrepancy may easily be explained by the fact that our calculations were carried out for the limit of small driving fields leaving the thermal occupation undisturbed whereas experimentally this limit was not achieved. With decreasing $T$ the
condition for a “weak” drive becomes more and more challenging
in the superconducting state as the energy relaxation time $\tau_{1}$
becomes very long. Thus resonant TSs may well be partially saturated at the given ultrasound intensity resulting in reduced absorption. This is supported by the observation of distorted pulse shapes indicating that steady state conditions are not achieved within the duration (typically $\unit[500]{ns}$) of a single ultrasound pulse. For a comparison, ultrasound echo experiments\cite{Weiss:PhononEchoesPdZr} on amorphous Pd$_{30}$Zr$_{70}$, the first direct measurements of energy relaxation and phase coherence times of a superconducting metallic glass, yield $\tau_1\approx30\mu$s and $\tau_2\approx5\mu$s at $\unit[20]{mK}$ for TS resonant with the ultrasonic frequency of $\unit[0.86]{GHz}$. New phonon echo experiments on $\text{Zr}_{59}\text{Ti}_{3}\text{Cu}_{20}\text{Ni}_{8}\text{Al}_{10}$ are presented in the supplemental material\citep{Suppl}.

The relaxation absorption in the superconducting state is characterised by a steep rise between 0.5 and 1\,K caused by the rapidly increasing quasiparticle density when approaching $T_{\text{c}}$. Measurement and calculation agree very well. In the normal conducting state the absorption depends only weakly on temperature.  As expected, the plateau of the relaxation absorption extends to low temperatures due to the slow variation of the dominating contribution of the electron bath to $\tau_1^{-1}$. Below 200\,mK, a clear however small increase of the absorption is observed indicating a resonant contribution. This behavior is remarkably well reproduced by our calculations. At first glance, one is led to expect a rather large resonant absorption since fast relaxation times would leave the TSs preferentially in their ground states except for extremely high ultrasonic intensities. However, caused by short dephasing times or, respectively, large line widths $h\tau_{2}^{-1}$ the resonant contribution to $\chi''$ is largely suppressed. This corresponds to the reduced slope of $\delta v(T)/v$.

\section{Conclusions}

This paper deals with low temperature ultrasound experiments on metallic glasses and their interpretation in terms of the elastic susceptibility of atomic two-level tunneling systems. It turns out that the standard tunneling model in its simplest version fails to describe the experiments properly, which is most evident for the temperature dependences of the sound velocity in the sample's superconducting state in comparison with its normal state. The simple analysis doesn't predict the experimentally observed crossing. 

In previous publications it was suggested that interaction with conduction electrons might in general reduce the density of states (DOS) of the tunneling systems thus leading to an overall smaller temperature variation of the sound velocity in the normal state compared to the superconducting state. This idea, however, is ruled out by the relaxation contribution to the attenuation $\tan\delta$ which obviously has the same absolute value in the normal state and the superconducting state below $T_{\rm c}$, however still in the plateau region (this is even better visible at frequencies between 201 and 731\,MHz in supplemental material\citep{Suppl}). In the present study, we show that another basic assumption of the treatment within the standard tunneling model is not fulfilled: Caused by their coupling to the bath of electronic excitations the tunneling systems no longer have line widths that are much smaller than their energy. This does not affect the relaxation contribution, however, considerably reduces the resonant contribution to the susceptibilities in the normal conducting metallic glass. 

Our numerical calculations employ the 'classical' approach for the resonant response of ensembles of two-level systems as given by eqs.~(\ref{eq:Chi-Resonant-Real}) and (\ref{eq:Chi-Resonant-Imag}) although this might be inadequate for extremely broad line widths caused by very high dephasing rates $\tau_{\phi}^{-1}$, which rather lead to localization and destruction of the quantum two-level nature. With $\tau_{\phi}^{-1}\propto T(T/T_0)^{-2/3}$, which emphasizes dephasing towards lower temperatures, we achieve very good overall agreement of our calculations with experiment, at least in the GHz frequency range. It is obvious that dephasing has to be consequently incorporated in such calculations. Usually, $\tau_{2}^{-1}\ll E/h$ is taken for granted. However, strong dephasing removes a considerable fraction of TS from contributing to $\chi_{\text{res}}$. A similar idea has been put forward for insulating glasses\citep{EnssHunklinger:IncoherentTunnelingVeryLowT}. It was suggested that strain mediated TS-TS interaction causes loss of coherence and a transition to incoherent tunneling, which finally leads to a reduced contribution to $\chi_{\text{res}}$.

Here, we have to ask: Which mechanism provides the temperature dependence of $\tau_{\phi}^{-1}$ in a metallic glass, only in the normal state? It is unlikely that electronic excitations are directly responsible for a dephasing rate variing other than linearly with temperature. Instead, we propose a dephasing mechanism that involves other TSs surrounding those which contribute to the resonant part of the susceptibility. We further propose that this TS-TS interaction is indirect and mediated by conduction electrons. Since the electron wave functions themselves become more coherent with decreasing temperature an increasing number of TSs will be within the volume defined by the respective phase coherence length and contribute to the dephasing of those resonant TSs. It is thought that both the TS-electron and the associated electron-TS interactions are longitudinal and combine to produce the suggested indirect TS-TS interaction. 

It is evident that the key for understanding the low temperature acoustic susceptibility of metallic glasses are the extremely high dephasing rates of TSs, the dynamics of which are largely determined by their interaction with conduction electrons. It is as well evident that the experiments at audio frequencies call for an additional explanation. As stated above, the straight $\log T$ dependence of $\delta v/v$ in the normal state at any frequency deserved particular attention. In any case, already \noun{Black}\footnote{\label{fn:Black-note} Fig.8.7 in the book chapter\cite{Black:TSinMetallicGlass}} expected a region of \emph{unusual behavior} when $E<h\tau_{2}^{\,-1}$.

Last but not least we like to note that the analysis of our measurements support quite generally the assumptions of the standard tunneling model including in particular the suggested distribution of relevant parameters. In this sense our work seems to constitute a good example of a 'smoking gun' experiment as was recently asked for by \noun{Leggett} and \noun{Vural}\cite{Leggett:TLSSmokingGun}. While the energy space is explored by the temperature from \unit[10]{mK} up to several Kelvin, the dynamics of these tunneling systems is measured at kHz and at GHz frequencies. Moreover, the dynamcis ($\tau_1$ and $\tau_2$) of TSs with same energy $E$ are vastly different whether the sample is superconducting or normal conducting.

\begin{acknowledgments}
We thank Alexander Prieschl and Markus Döttling as well as Sebastian Craft for contributing to the measurements at 1.4 GHz and at audio frequencies, respectively. We thank Wolfgang Wernsdorfer for his generous financial support. GW, in particular, is grateful to Siegfried Hunklinger for 40 years of continuous interest in this work - Siggi, happy birthday to your 80th.

AM and TV conducted the measurements, SM prepared the ZnO transducer for ultrasonic measurements, UK and SS provided the metallic glass samples, AS contributed to the discussion, AM performed the numerical calculations, and wrote the manuscript together with GW.
\end{acknowledgments}

\bibliographystyle{natbib}

\begin{thebibliography}{99} 


\bibitem{HunklingerArnold:UltrasonicGlass}
S. Hunklinger and W. Arnold.
\newblock {Ultrasonic Properties of Glasses at Low Temperatures}.
\newblock In W.P. Mason and R.N. Thurston, editors, {\em {Physical Acoustics}}, volume 12, page 155--215. Academic Press, 1976.

\bibitem{Phillips:TLScollection}
W.A. Phillips.
\newblock {Two-level states in glasses}.
\newblock {\em Reports on Progress in Physics}, 50(12):1657, 1987.

\bibitem{Phillips1981:AmorphousSolids}
W.A. Phillips.
\newblock {\em {Amorphous Solids}}, volume 24 of {\em {Topics in Current
  Physics}}.
\newblock Springer-Verlag Berlin Heidelberg, 1981.

\bibitem{ZellerPohl:TunnelingModell}
R.C. Zeller and R.O. Pohl.
\newblock {Thermal Conductivity and Specific Heat of Noncrystalline Solids}.
\newblock {\em Phys. Rev. B}, 4:2029--2041, sep 1971.

\bibitem{Anderson:StandardTunnelingModel}
P.W. Anderson, B.I. Halperin, and M.C. Varma.
\newblock {Anomalous low-temperature thermal properties of glasses and spin
  glasses}.
\newblock {\em Philosophical Magazine}, 25(1):1--9, 1972.

\bibitem{Phillips1972:TLS}
W.A. Phillips.
\newblock {Tunneling states in amorphous solids}.
\newblock {\em Journal of Low Temperature Physics}, 7(3):351--360, 1972.

\bibitem{Esquinazi:VibratingReedSuperconductingMetGlass}
P. Esquinazi, H.M. Ritter, H. Neckel, G. Weiss, and S. Hunklinger.
\newblock {Acoustic experiments on amorphous Pd$_{30}$Zr$_{70}$ and
  Cu$_{30}$Zr$_{70}$ in the superconducting and the normal state}.
\newblock {\em Zeitschrift für Physik B Condensed Matter}, 64(1):81--93,
  1986.

\bibitem{Neckel:TunnelingModelIncomplete}
H. Neckel, P. Esquinazi, G. Weiss, and S. Hunklinger.
\newblock {The tunneling model — Incomplete for amorphous metals}.
\newblock {\em Solid State Communications}, 57(3):151--154, 1986.

\bibitem{Jaeckle:UltrasonicAttenuationGlasses}
J. Jäckle.
\newblock {On the ultrasonic attenuation in glasses at low temperatures}.
\newblock {\em Zeitschrift für Physik}, 257(3):212--223, 1972.

\bibitem{Black:TSinMetallicGlass}
J.L. Black.
\newblock {Low-energy excitations in metallic glasses}.
\newblock In Hans-Joachim Güntherodt and Hans Beck, editors, {\em {Glassy
  Metals I}}, page 167--190. Springer Berlin Heidelberg, 1981.

\bibitem{Golding:RelaxationCondElectrons}
B. Golding, J.E. Graebner, A.B. Kane, and J.L. Black.
\newblock {Relaxation of Tunneling Systems by Conduction Electrons in a
  Metallic Glass}.
\newblock {\em Phys. Rev. Lett.}, 41:1487--1491, Nov 1978.

\bibitem{BezuglyiZherlitsyn:ElectronRenormalizationBMG}
E.V. Bezuglyi, A.L. Gaiduk, V.D. Fil, S. Zherlitsyn, W.L. Johnson,
  G. Bruls, B. Lüthi, and B. Wolf.
\newblock {Electron renormalization of sound interaction with two-level
  systems in superconducting metallic glasses}.
\newblock {\em Phys. Rev. B}, 62(10):6656--6664, Sep 2000.

\bibitem{Black:RenormalizationTheoryTLSinMetallicGlass}
J.L. Black, K. Vladár, and A. Zawadowski.
\newblock {Renormalization-group theory for the commutative model of tunneling
  states in metallic glasses}.
\newblock {\em Phys. Rev. B}, 26:1559--1568, Aug 1982.

\bibitem{KaganProkov:Renormalization}
Yu. Kagan and N.V. Prokof'ev.
\newblock {Acoustic properties of metallic and superconducting glasses}.
\newblock {\em Solid State Communications}, 65(11):1385--1389, 1988.

\bibitem{Note1}
Named after \protect \textsc {G. Wenzel}, \protect \textsc {H. A. Kramers} and
  \protect \textsc {L. Brillouin}.

\bibitem{Maynard1976:MultiPhonon}
R. Maynard.
\newblock {\em {Low Energy Phonons in Amorphous Materials}}, page 115--122.
\newblock Springer US, Boston, MA, 1976.

\bibitem{Black:SuperconductingMetallicGlassTLS}
J.L. Black and P. Fulde.
\newblock {Influence of the Superconducting State upon the Low-Temperature
  Properties of Metallic Glasses}.
\newblock {\em Phys. Rev. Lett.}, 43:453--456, Aug 1979.

\bibitem{WangsnessBloch:NuclearInduction}
R.K. Wangsness and F. Bloch.
\newblock {The Dynamical Theory of Nuclear Induction}
\newblock {\em Phys. Rev.}, 89:728-739, Feb. 1953.

\bibitem{Redfield:RelaxationNuclearSpin}
A.G. Redfield.
\newblock{On the Theory of Relaxation Processes}
\newblock {\em IBM Journal of Research and Development}, 1:19-31, 1957.

\bibitem{Shnirman:DissipationQubits}
Yu. Makhlin, G. Schön, and A. Shnirman.
\newblock {Dissipation in Josephson qubits}.
\newblock In R. Fazio, V.F. Gantmakher, and Y. Imry, editors, {\em {New
  Directions in Mesoscopic Physics (Towards Nanoscience)}}, page 197--224,
  Dordrecht, 2003. Springer Netherlands.
\newblock { arXiv:cond-mat/0309049 }.

\bibitem{Note2}
A $(\varepsilon /E)^{2}$ dependence of $\tau _{\phi }^{\protect
  \tmspace +\thinmuskip {.1667em}-1}$, originating from the TSs' dipolar
  interactions, has been observed in dielectric echo experiments on individual
  TSs in the oxide films of Josephson junction qubits \protect \citep
  {Lisenfeld:DecoherenceSpectroscopySREP}.

\bibitem{Lisenfeld:DecoherenceSpectroscopySREP}
J. Lisenfeld, A. Bilmes, S. Matityahu, S. Zanker,
  M. Marthaler, M. Schechter, G. Schön, A. Shnirman, G.
  Weiss, and A.V. Ustinov.
\newblock {Decoherence spectroscopy with individual two-level tunneling
  defects}.
\newblock {\em Scientific Reports}, 6(23786):2045--2322, Mar 2016.
\newblock { arXiv:1601.03213 } 

\bibitem{Kuehn:ZrTiCuNiAl}
U. Kühn.
\newblock {\em {Strukturelle und mechanische Charakterisierung von
  vielkomponentigen amorphen, teilamorphen und kristallinen
  Zirkon--Basislegierungen}}.
\newblock PhD thesis, Technische Universität Dresden, 2004.

\bibitem{Berry:IBMVibratingReed}
B.S. Berry and W.C. Pritchet.
\newblock {Vibrating Reed Internal Friction Apparatus for Films and Foils}.
\newblock {\em IBM Journal of Research and Development}, 19(4):334--343, Jul
  1975.

\bibitem{Classen:LowFreqAcousticDielectricGlass}
J. Classen, C. Enss, C. Bechinger, G. Weiss, and S. Hunklinger.
\newblock {Low frequency acoustic and dielectric measurements on glasses}.
\newblock {\em Annalen der Physik}, 506(5):315--335, 1994.

\bibitem{Luethi:Physical_Acoustics_in_the_solid_State}
B. Lüthi.
\newblock {\em {Physical Acoustics in the Solid State}}.
\newblock Springer, Berlin, Heidelberg, New York, springer series solid-state
  science 148 edition, 2005.

\bibitem{Doettling:VibratingReed}
M. Döttling.
\newblock {Niederfrequente akustische Messungen an amorphem Zr$_{59}$ Ti$_{3}$
  Cu$_{20}$ Ni$_{8}$ Al$_{10}$ }.
\newblock Bachelor's thesis, Karlsruher Institut für Technologie (KIT), Nov 2016, unpublished.

\bibitem{Seiler:MetGlasSound}
A. Seiler.
\newblock {\em {Einfluss der Leitungselektronen auf die Dynamik atomarer
  Tunnelsysteme in ungeordneten Festkörpern: Relaxationsprozesse in
  metallischen Gläsern und ungeordneten dünnen Aluminiumoxid-Schichten}}.
\newblock PhD thesis, Karlsruher Institut für Technologie (KIT), 2019.

\bibitem{Note3}
Lapping was done by \protect \textsc {Stähli-Läpptechnik}, Weil im
  Schönbuch, Germany.

\bibitem{Raychaudhuri:ElasticPropGlasses}
A.K. Raychaudhuri and S. Hunklinger.
\newblock {Low frequency elastic properties of glasses at low
  temperatures—implications on the tunneling model}.
\newblock {\em Zeitschrift für Physik B Condensed Matter}, 57(2):113--125,
  1984.

\bibitem{SchneiderSun:OxidationJMR1996}
X. Sun, S. Schneider, U. Geyer, W.L. Johnson, and M.A. Nicolet.
\newblock {Oxidation and crystallization of an amorphous Zr$_{60}$Al$_{15}$Ni$_{25}$ alloy}.
\newblock {\em Journal of materials research}, 11(11):2738--2743, 1996.

\bibitem{Adler:VibReedCJP2011}
data include those by C. Adler, W. Mohr, T. Peichl, and G. Weiss.
\newblock {Low Temperature Vibrating Reed Measurements of Zr Based Metallic
  Glasses}.
\newblock {\em Chinese Journal of Physics}, 49(1):429, Feb 2011.

\bibitem{UniversalAcousticProps}
J.F. Berret and M. Meißner.
\newblock {How universal are the low temperature acoustic properties of
  glasses?}
\newblock {\em Zeitschrift für Physik B Condensed Matter}, 70(1):65--72,
  1988.

\bibitem{Doussineau:FlourideGlass}
P. Doussineau, M. Matecki, and W. Schön.
\newblock {Connection between the low temperature acoustic properties and the
  glass transition temperature of fluoride glasses}.
\newblock {\em J. Phys. France}, 44(1):101--107, 1983.

\bibitem{Burin:CollectiveExcitationGlass}
A.L. Burin and Yu. Kagan.
\newblock {Low-energy collective excitations in glasses. New relaxation
  mechanism for ultralow temperatures}.
\newblock {\em Journal of Experimental and Theoretical Physics}, 79(2):347,
  1994.

\bibitem{Burin:DipoleGapRelaxation}
A.L. Burin.
\newblock {Dipole gap eﬀects in low energy excitation spectrum of amorphous
  solids. Theory for dielectric relaxation}.
\newblock {\em Journal of Low Temperature Physics}, 100(3):309--337, 1995.

\bibitem{Rothfuss:ThermCondBMG}
D.S. Rothfuss, U. Kühn, A. Reiser, A. Fleischmann, and C. Enss.
\newblock {Thermal Conductivity of Superconducting Bulk Metallic Glasses at
  Very Low Temperatures}.
\newblock {\em Chinese Journal of Physics}, 49(384), 2011.

\bibitem{Rothfuss:ThermalPropBMG}
D.S. Rothfuss.
\newblock {\em {Thermal properties of superconducting bulk metallic glasses at
  ultralow temperatures}}.
\newblock PhD thesis, Universität Heidelberg, Nov 2013.

\bibitem{Weiss:PhononEchoesPdZr}
G. Weiss and B. Golding.
\newblock {Phonon Echoes in Superconducting Amorphous Pd$_{30}$ Zr$_{70}$ }.
\newblock {\em Phys. Rev. Lett.}, 60:2547--2550, Jun 1988.

\bibitem{EnssHunklinger:IncoherentTunnelingVeryLowT}
C. Enss and S. Hunklinger.
\newblock {Incoherent Tunneling in Glasses at Very Low Temperatures}.
\newblock {\em Phys. Rev. Lett.}, 79:2831--2834, Oct 1997.

\bibitem{Leggett:TLSSmokingGun}
A.J. Leggett and D.C. Vural.
\newblock {“Tunneling Two-Level Systems” Model of the Low-Temperature
  Properties of Glasses: Are “Smoking-Gun” Tests Possible?}
\newblock {\em The Journal of Physical Chemistry B}, 117(42):12966--12971,
  2013.
\newblock { arXiv:1310.3387 }

\bibitem{Note4}
\label {fn:Black-note} Fig.~8.7 in the book chapter\cite
  {Black:TSinMetallicGlass}.

\bibitem{Suppl}
Supplemental material below.

\end{thebibliography}

\begin{thebibliography}{999}


\bibitem[S1]{Suppl-Schneider:BulkMetallicGlasses}
S. Schneider.
\newblock {Bulk metallic glasses}.
\newblock {\em Journal of Physics: Condensed Matter}, 13(34):7723--7736, Aug
  2001.

\bibitem[S2]{Suppl-SchneiderSun:OxidationJMR1996}
X. Sun, S. Schneider, U. Geyer, W.L. Johnson, and M.A. Nicolet.
\newblock {Oxidation and crystallization of an amorphous Zr$_{60}$Al$_{15}$Ni$_{25}$ alloy}.
\newblock {\em Journal of materials research}, 11(11):2738--2743, 1996.

\bibitem[S3]{Suppl-Phillips1972:TLS}
W.A. Phillips.
\newblock {Tunneling states in amorphous solids}.
\newblock {\em Journal of Low Temperature Physics}, 7(3):351--360, 1972.

\bibitem[S4]{Suppl-Anderson:StandardTunnelingModel}
P.W. Anderson, B.I. Halperin, and M.C. Varma.
\newblock {Anomalous low-temperature thermal properties of glasses and spin
  glasses}.
\newblock {\em Philosophical Magazine}, 25(1):1--9, 1972.

\bibitem[S5]{Suppl-Golding:PhononEchoesGlass}
B. Golding and J.E. Graebner.
\newblock {Phonon Echoes in Glass}.
\newblock {\em Phys. Rev. Lett.}, 37:852--855, Sep 1976.

\bibitem[S6]{Suppl-GoldingHunklinger:PhotonEchoSilica}
B. Golding, M.v. Schickfus, S. Hunklinger, and K. Dransfeld.
\newblock {Intrinsic Electric Dipole Moment of Tunneling Systems in Silica Glasses}.
newblock {\em Phys. Rev. Lett.}, 43:1817-1821, Dec 1979.

\bibitem[S7]{Suppl-Golding-Graebner:inPhillips1981}
B. Golding and J.E. Graebner. in W.A. Phillips (ed.)
\newblock {\em {Amorphous Solids}}, volume~24 of {\em {Topics in Current
  Physics}}, p.109-134, 
\newblock Springer-Verlag Berlin Heidelberg, 1981. 

\bibitem[S8]{Suppl-Weiss:PhononEchoesPdZr}
G. Weiss and B. Golding.
\newblock {Phonon Echoes in Superconducting Amorphous Pd$_{30}$ Zr$_{70}$ }.
\newblock {\em Phys. Rev. Lett.}, 60:2547--2550, Jun 1988.

\end{thebibliography}



\cleardoublepage{}

\section{Supplemental}

These supplemental parts contain i) some more information on the possible energy dependence of the density of states (DOS) of two level tunneling states (TS) in the glassy metal $\text{Zr}_{59}\text{Ti}_{3}\text{Cu}_{20}\text{Ni}_{8}\text{Al}_{10}$, ii) further measurements on the ultrasound absorption at frequencies from 300\,MHz to 700\,MHz, present iii) phonon echo experiments of the Hahn echo type measuring $\tau_2$ times of TSs resonant with ultrasound of 1 and 1.3\,GHz, and show iv) vibrating reed measurements of the splat cooled metallic glass Zr$_{46.8}$Ti$_{8.2}$Cu$_{7.5}$Ni$_{10}$Be$_{27.5}$ (Vitreloy 4)\cite{Suppl-Schneider:BulkMetallicGlasses,Suppl-SchneiderSun:OxidationJMR1996}.

\subsection{DOS of TS energy dependent?}

\begin{figure}
\includegraphics{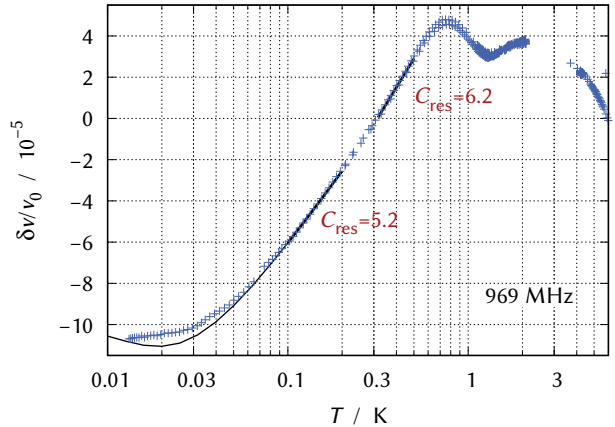}
\caption{\label{slopeIncrease}Temperature dependence of the transverse sound velocity measured at 969 MHz. Experimental data are the same as of Fig.\,1 of the main article. Solid lines represent numerical calculations plotted for two confined temperature ranges. $C_{\text res}$ values ('$\log T$ slopes') indicate a possible reduction of the TSs' DOS at lower engergies by $\approx18\%$. The calulation for the low temperature range shows the expected minimum at this measuring frequency. Obviously, the sample is thermally decoupled below $T\sim70$\,mK. }
\end{figure}

As mentioned in the main text along with discussing temperature dependencies of the sound velocity $\delta v(T)/v$ at various frequencies a change in the '$\log T$ slope' may imply a variation of the DOS of TS. Fig.\,\ref{slopeIncrease} shows again the $\delta v(T)/v$ measurements at 969\,MHz, and for the discussion slopes are added in two temperature regions, around 500 and 150\,mK. This suggests a small decrease of the TSs' DOS for respective energies. However, this effect should be taken with caution. First of all, we cannot be sure that the sample is thermally equilibrated with the thermometer. And second as already pointed out, measurements of the thermal conductivity of a closely related metallic glass, superconducting as well, don't support any deviation from a constant density of states of TS. Both quantities, sound velocity variations in the superconducting state as well as thermal conductivity measure predominantly TSs with almost symmetric double well potentials.

\begin{figure}
\includegraphics{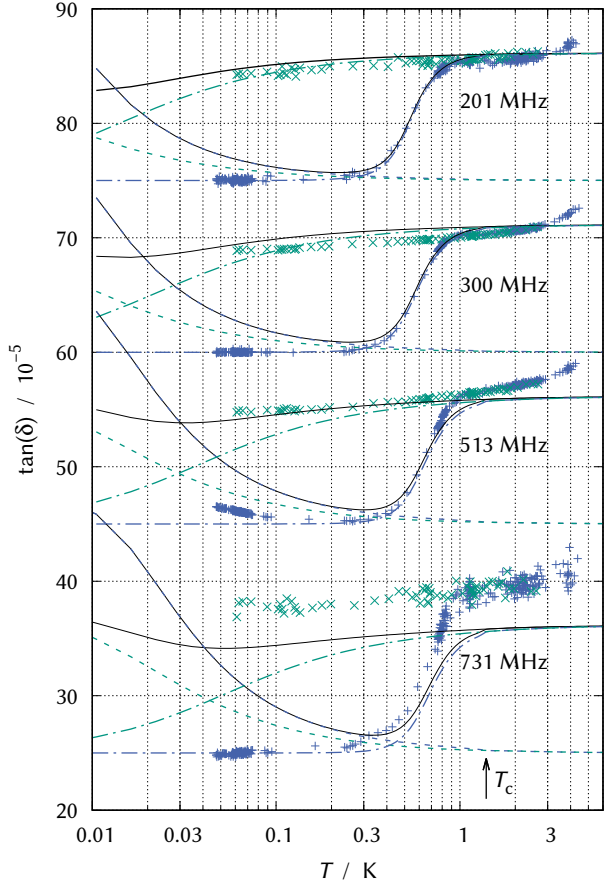}
\caption{\label{loss-tangent_300M-1G6}Temperature dependence of the transverse ultrasound absorption at frequencies from 201 to 731\,MHz. Lines show calculations: dash-dotted for relaxation contributions, short dashed for resonant contributions, green for normal, and blue for superconducting below $T_{\text c}$. Parameters have the same numerical values as in fig.\,6 of the main article. The measurement at 731\,MHz suffered from considerable noise caused by suboptimal transducer thickness and frequency band of the RF-components.}
\end{figure}

Ultrasonic absorption as shown in fig.\,6 of the main article and again most clearly in fig.~\ref{loss-tangent_300M-1G6} exhibts a plateau extending in the normal conducting state to temperatures as low as 60\,mK. For the lower frequencies this absorption is mainly caused by the relaxation mechanism of TSs with a wide distribution of asymmetry energies and the plateau indicates a constant density of states with distributions of asymmetry energies and tunneling parameters as proposed in the original tunneling model\cite{Suppl-Phillips1972:TLS,Suppl-Anderson:StandardTunnelingModel}. The decreasing relaxation absorption towards lower temperature is partially balanced by resonant absorption, particularly for the measurements at higher frequencies. For better illustration, the two absorption mechanisms as calculated for both normal and superconducting states are separately plotted in fig.\,\ref{loss-tangent_300M-1G6}. As explained in the main article, the resonant absorption in the normal metal is strongly reduced due to the TSs' short coherence time whereas in the superconducting sample it is partially or completely saturated.  

As discussed in the main article, we finally like to stress again that the straight $\log T$ of $\delta v/v$ in the normal state poses quite hard constraints for any model that generates deviating distribution functions and DOS of TS. It requires a delicate balance of resonant and relaxation contributions of TSs with quite different ratios of tunneling energy and asymmetry energy. In general terms this is achieved by the assumptions of the Tunneling Model. 

\subsection{Phonon echoes}

Quite similar to other two-level systems, e.g. nuclear or electron spins or electronic transitions in atoms and molecules, atomic-tunneling systems (TS) in solids may be excited and driven by microwave pulses to generate echoes. Equipped with usually both elastic and electric dipole moments, TS may respond with phononic\cite{Suppl-Golding:PhononEchoesGlass} as well as photonic\cite{Suppl-GoldingHunklinger:PhotonEchoSilica} echoes. The latter one actually provided details of the microscopic nature of TS\cite{Suppl-GoldingHunklinger:PhotonEchoSilica}. Peculiar to phonon echoes is that they propagate co-linearly behind their generating ultrasound pulses collecting energy from excited TSs along the propagation path. 

Details of the physics of phonon echoes in glasses, two-pulse echoes to determine transverse relaxation times $\tau_2$ of a spectral range of TS defined by the spectrum of the driving pulses, and three-pulse echoes for $\tau_1$, the energy relaxation time to establish thermal population of the two levels, are well described in the literature\cite{Suppl-Golding:PhononEchoesGlass,Suppl-Golding-Graebner:inPhillips1981}. In dielectric glasses and in superconducting glasses well below the transition temperature\cite{Suppl-Weiss:PhononEchoesPdZr} $\tau_1$ is determined by one-phonon processes, i.e. emission or absorption of a resonant thermal phonon of the bath.  

In the present study of the bulk metallic glass $\text{Zr}_{59}\text{Ti}_{3}\text{Cu}_{20}\text{Ni}_{8}\text{Al}_{10}$ two-pulse echoes were employed to measure $\tau_2$ at 11\,mK for two frequencies, 1002 and 1367\,MHz. Ultrasound pulses were generated at one side of the sample and detected in transmission by another ZnO transducer on the opposite side. Since the first ultrasound pulse is stronger reduced by resonant absorption than the second pulse it is not obvious were along the propagation path the optimal condition of a $\pi/2 - \pi$ -- pulse sequence occurs. The received signals always contain the full 'history' of non-linear absorption as well as generation of echoes of the propagating ultrasound pulses. 

\begin{figure}
\includegraphics{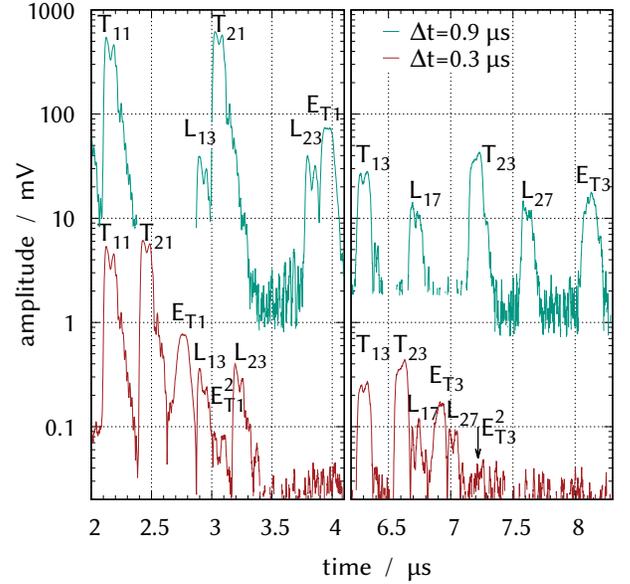}
\caption{\label{Echo_Skizze_2} Received signals on log-scale of transverse (\textsf{T}) and longitudinal (\textsf{L}) 1.002\,GHz ultrasound pulses of 100\,ns duration, and generated two-pulse echoes (\textsf{E}). The first pulse is launched at time $t=0$ followed by the second pulse after delay times $\Delta t= 0.3\,\mu$s (lower signal traces in red) or $\Delta t= 0.9\,\mu$s (upper signal traces in green). First indices 1 or 2 at the received signals \textsf{T} and \textsf{L} refer to these two pulses. The respective second indices at \textsf{T} and \textsf{L} label the number of passes travelled by the pulses. Traces are vertically shifted for clarity of the graph and low amplitude noise is cutoff. }
\end{figure}

In most of our experiments (as in the original work\cite{Suppl-Golding:PhononEchoesGlass}) two identical acoustic pulses were transmitted. Signals from regular ultrasound pulses and echoes were detected by the receiving transducer at times $t=T_{\text t,l}, 3T_{\text t,l}, 5T_{\text t,l}, ...$, i.e. after odd multiples of the time $T_{\text t,l}=L/v_{\text t,l}$ given by the sample length $L$ divided by the (transverse or longitudinal) speed of sound $v_{\text t,l}$. Figure\,\ref{Echo_Skizze_2} shows received signals after $t=T_{\text t}$ in the left panel and after $t=3T_{\text t}$ in the right panel. Caused by a small inhomogeneity of the sample material and related variation of phase velocity the integrated signal across the transducer area suffered from interferences and therefore the shapes of all detected pulses were distorted. Resolving and identifying all signal pulses were further impeded by the crystalline structure of the ZnO transducers which generated simultaneously both transverse and longitudinal ultrasound. 

In the left panel of fig.\,\ref{Echo_Skizze_2} clear (transverse) phonon echo signals \textsf{E}$_{\textsf{T1}}$ are observed at twice the time delay $\Delta t$ from the first ultrasound pulse \textsf{T}$_{11}$ to the second one \textsf{T}$_{21}$. Pulses of longitudinal ultrasound \textsf{L}$_{13}$ and \textsf{L}$_{23}$ are seen as well, no longitudinal echoes could be observed. However, shortly after \textsf{L}$_{13}$ a secondary echo \textsf{E}$^2_{\textsf{T1}}$ is generated by \textsf{T}$_{21}$ and the primary echo \textsf{E}$_{\textsf{T1}}$. The amplitude of \textsf{E}$^2_{\textsf{T1}}$ is another factor 10 smaller than \textsf{E}$_{\textsf{T1}}$ which in turn is smaller by factor 10 than the sound pulse \textsf{T}$_{21}$.

In the right panel of fig.\,\ref{Echo_Skizze_2} all signals can be found again, at a later time, when the transverse ultrasound pulses have travelled three times the sample length; longitudinal sound pulses 7 times. It is quite evident now, that the first ultrasound pulse \textsf{T}$_{13}$ is stronger attenuated than the second one \textsf{T}$_{23}$ and that the echo amplitude \textsf{E}$_{\textsf{T3}}$ is grown with respect to their generating ultrasound pulses. The secondary echo \textsf{E}$^2_{\textsf{T3}}$ appears just above noise level. It is also evident from the upper (green) signals that the two longitudinal sound pulses keep the same relative height even after 7 passes through the sample. This may be interpreted as a weaker coupling to tunneling systems compared to transverse sound, so that the first pulse could not noticeably change the two-level occupation numbers. 

\begin{figure}
\includegraphics{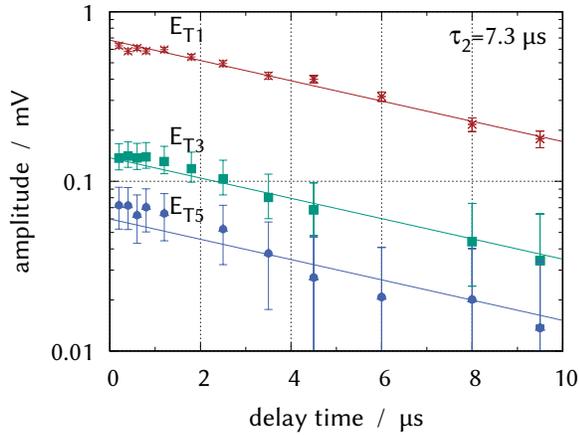}
\caption{\label{Echo_Amplitude_1002MHz}Amplitudes of phonon echoes \textsf{E}$_{\textsf{T1}}$, \textsf{E}$_{\textsf{T3}}$, and \textsf{E}$_{\textsf{T5}}$ at 1.002\,GHz in dependence of the delay time $\Delta t$.}
\end{figure}
 
Figure\,\ref{Echo_Amplitude_1002MHz} shows signals of ultrasound echoes after 1, 3, and 5 passes through the sample the heights of which are exponentially decreasing with increasing delay between the two generating pulses. A fit with \textsf{E}$\,\propto\exp{-t/\tau_2}$ yields a transverse relaxation time $\tau_2=7.3\mu$s for TS within the spectral range of a 100\,ns pulse at 1\,GHz and at 11\,mK. Analog experiments at 1.367\,GHz showed an overall similar picture, however with $\tau_2=5.2\mu$s indicating that $\tau_2$ is roughly proportional to the the TS energy -- a major result of these phonon echo experiments.

\subsection{Vibrating Reed Measurements of Zr$_{46.8}$Ti$_{8.2}$Cu$_{7.5}$Ni$_{10}$Be$_{27.5}$}

\begin{figure}
    \includegraphics{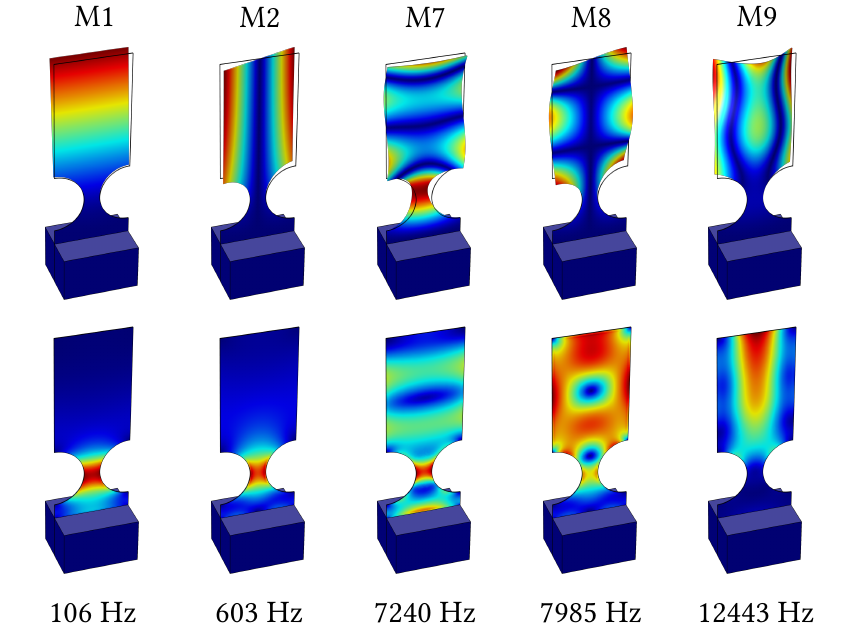}
    \caption{\label{some_reed-modes} Results of finite element calculations show various modes of the specially shaped sample. Electrodes for electrostatic drive and detection are mounted slightly off-center. Different modes are not equally well detectable. Modes are shown form left to right with increasing eigenfrequencies as noted. Top figures represent displacements and bottom figures corresponding stress distributions in color code from blue (small) to red (large).}
\end{figure}

These measurements are briefly described since some of the results are used in fig.\,4 of the main article to support the $T^3$ dependence of the TSs' relaxation rate in the superconduction state at lowest temperatures. The splat cooled sample material was tailored to a shape that allowed for a series of vibrational resonances, both flexural and torsional modes, at frequencies between 100\,Hz and 12\,kHz. A few examples are shown in fig.\,\ref{some_reed-modes} as color coded images of displacements and respective stress fields. It is important to note that deformation at the clamping is negligible or small for all modes and clamping losses may be disregarded. 

\begin{figure}
\includegraphics{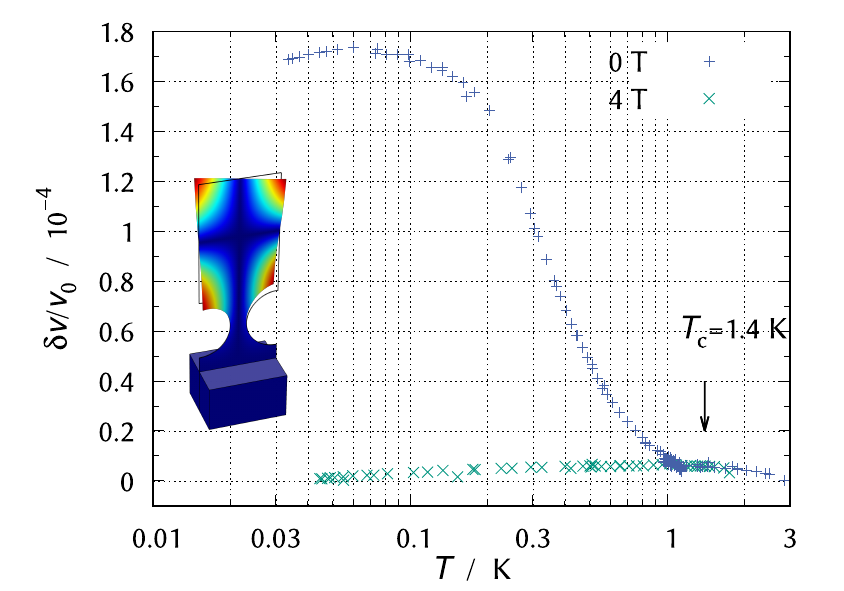}
\caption{\label{splat_BMG_dv} Temperature dependence of the sound velocity of splat-quenched glassy Zr$_{46.8}$Ti$_{8.2}$Cu$_{7.5}$Ni$_{10}$Be$_{27.5}$, measured with mode 4 -- a flexural mode -- at 2815 Hz. Material is superconducting below 1.4\,K, a magnetic field of 4\,T keeps it normal conducting. }
\end{figure}

Modes at lowest frequencies are a simple nodding or rotation (see modes 1 and 2 in fig.\,\ref{some_reed-modes}) of the large plate with restoring forces produced by bending or twisting the narrow strip. Mode 4 is an antisymmetric flexural mode characterised by a blue cross (see fig.\,\ref{splat_BMG_dv}) for smallest displacements. Higher modes vibrate with displacement node lines in the shape of a double cross or horns (modes 8 and 9). 

\begin{figure}
\includegraphics{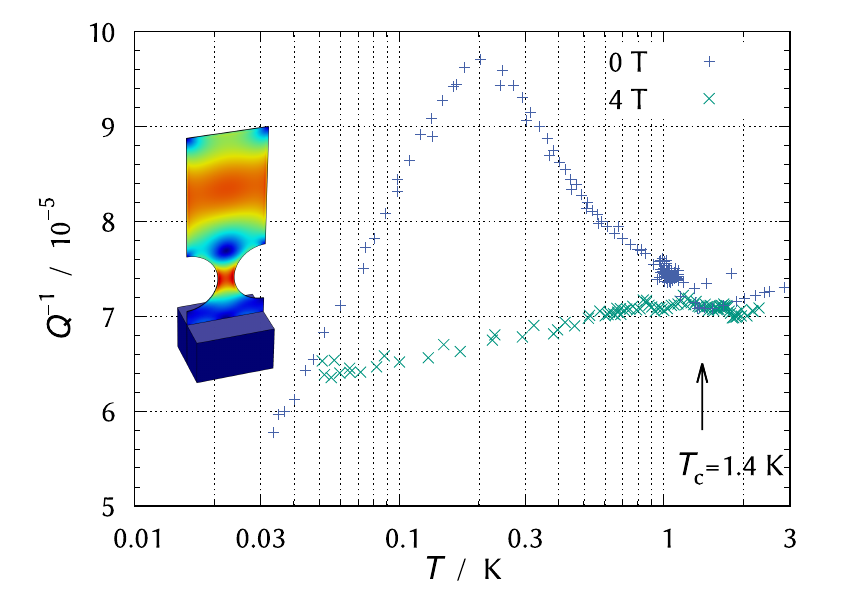}
\caption{\label{splat_BMG_Q} Absorption $Q^{-1}$ of Zr$_{46.8}$Ti$_{8.2}$Cu$_{7.5}$Ni$_{10}$Be$_{27.5}$ and stress distribution of the employed flexural mode corresponding to fig.\,\ref{splat_BMG_dv}.}
\end{figure}

As an example, fig.\,\ref{splat_BMG_dv} shows the temperature dependence of the sound velocity of mode 4 at 2815 Hz in the superconducting and in the normal states. The general behaviour resembles that of other superconducting metallic glasses as presented and discussed in the main part of this paper. Of particular importance here is the maximum at about 60 mK in the superconducting sample marking the onset of relaxation contributions to the susceptibility. The temperature of this onset shifts with frequency as plotted in fig.\,4 of the main part of this paper and confirms that the relaxation of TS is caused by one-phonon processes at temperatures well below the superconducting transition. 

It is worth noting that the one-phonon rate is dominated by the spectral density of transverse thermal phonons. Therefore all modes of the sample can be used for this analysis, irrespective of the mode being flexural of torsional. 

\begin{figure}
\includegraphics{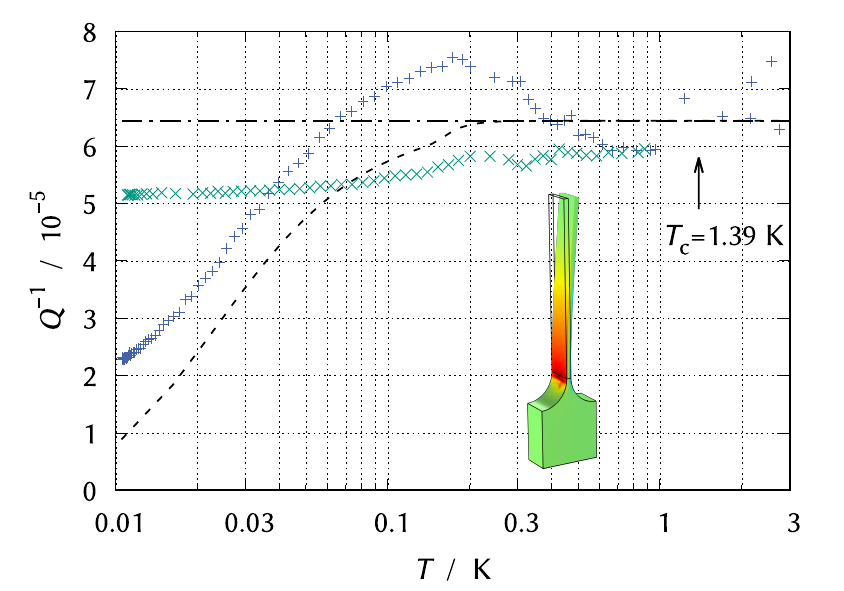}
\caption{\label{DresdenGlas_Q} Absorption of the bulk metallic glass $\text{Zr}_{59}\text{Ti}_{3}\text{Cu}_{20}\text{Ni}_{8}\text{Al}_{10}$ corresponding to the sound velocity measurement of fig.\,5 in the main text. Dashed lines are calculations with parameters according to those data.}
\end{figure}

Figure \ref{splat_BMG_Q} finally shows absorption data corresponding to fig.\,\ref{splat_BMG_dv}. It exhibits a pronounced maximum in the superconducting state, about 40\% higher than the absorption in the normal state. The absorption in the normal state is gradually increasing by about 10\% between 50\,mK and 1\,K. Similar behavior is found for other metallic glasses as shown e.g. in the last figure \ref{DresdenGlas_Q}. However, this is not predicted by the standard tunneling model as already discussed in the main part of the publication. 

The relaxation mechanism, particularly at low frequencies involves and tests the contribution of a class of TS with extremely small tunneling energy $\varDelta$ compared to their energy splitting $E$. Note, that only a perfect distribution function $P\propto 1/\varDelta$ generates a perfect plateau of the relaxation contribution to $Q^{-1}$. Thus, quite different sections of this distribution are involved whether TS interact only with thermal phonons or additionally with conduction electrons or quasiparticles of the superconductor to achieve their equilibrium population. As pointed out already, minute modifications of the $\propto 1/\varDelta$ distribution may contribute to solve this discrepancy.


\end{document}